\documentclass[3p, onecolumn, numbers]{elsarticle} 
\usepackage{hyperref,color}
\usepackage{todonotes}
\usepackage{ulem}
\usepackage{amssymb}
\usepackage{booktabs}
\usepackage{chemformula}
\usepackage{gensymb}
\usepackage{comment}
\usepackage{svg}
\setchemformula{radical-radius=1pt}
\RequirePackage{xspace}
\DeclareFontFamily{OT1}{pzc}{}
\DeclareFontShape{OT1}{pzc}{m}{it}{<-> s * [1.200] pzcmi7t}{}
\DeclareMathAlphabet{\mathpzc}{OT1}{pzc}{m}{it}
\usepackage{amsfonts}
\usepackage{textcomp}
\journal{Nuclear Instruments and Methods A}
\usepackage{graphicx}
\usepackage{caption}
\usepackage{subcaption}
\usepackage{siunitx}
\usepackage[utf8]{inputenc}

\usepackage{multirow}

\begin{document}

\begin{frontmatter}

\title{Performance Evaluation of Straw Tubes with Muon Beams at CERN}

\author[umich]{Linnuo Zhang\corref{cor}}
\cortext[cor]{Corresponding author}
\ead{linnuo@umich.edu}
\author[umich]{Chihao Li}
\author[umich]{Jiajin Ge}
\author[tufts]{Tatiana Azaryan}
\author[dubna]{Vitalii Bautin}
\author[dubna]{Artem Chukanov}
\author[umich]{Tiesheng Dai}
\author[kazakhstan,dubna]{Temur Enik}

\author[umich]{Liang Guan}
\author[umich]{Yuxiang Guo}
\author[umich]{Jiahao Hu}
\author[kazakhstan,florida]{Ekaterina Kuznetsova}

\author[umich]{Hui-Chi Lin}
\author[umich]{Jianming Qian}
\author[cern]{Andre Rummler}
\author[umich]{Emmett Salzer}
\author[NRC]{Dmitry Sosnov}
\author[umich]{Can Suslu}
\author[umich]{Curtis Weaverdyck}
\author[umich]{Frances Wharton}
\author[umich]{Ruslan Yakubovych}
\author[umich]{Bing Zhou}
\author[umich]{Jessaly Zhu}
\author[umich]{Junjie Zhu}

\affiliation[umich]{organization={University of Michigan}, city={Ann Arbor}, state={MI}, country={USA}}
\affiliation[tufts]{organization={Tufts University}, city={Medford}, state={MA}, country={USA}}
\affiliation[dubna]{organization={Joint Institute for Nuclear Research}, city={Dubna}, country={Russia}}
\affiliation[kazakhstan]{organization={Institute of Nuclear Physics}, city={Almaty}, country={Kazakhstan}}
\affiliation[florida]{organization={University of Florida}, city={Gainesville}, state={FL}, country={USA}}
\affiliation[cern]{organization={CERN, European Organization for Nuclear Research}, city={Geneva}, country={Switzerland}}
\affiliation[NRC]{organization={NRC Kurchatov Institute PNPI}, city={Gatchina}, country={Russia}}

\begin{abstract}
We present results from two test beam campaigns that investigate the performance of straw tube detectors as potential candidates for an FCC-ee straw tracker. These studies were carried out at CERN using 150 GeV muon beams. Dedicated algorithms were developed to determine both single tube spatial resolution for the primary coordinate in the $r-\phi$ plane and spatial resolution for the secondary coordinate along the tube direction within a straw chamber. Detection efficiency was also evaluated as a function of the extrapolated hit position for each tube. Both datasets showed consistent results for spatial resolutions and efficiency.  The average single-tube spatial resolution in the primary  $r-\phi$ coordinate is approximately 110 {\micro m},  while the average spatial resolution in the secondary coordinate along the tube direction is approximately 2 mm. Most tubes show a single-tube detection efficiency above 98\%.

\end{abstract}

\begin{keyword}
tracker\sep straw\sep spatial resolution\sep detection efficiency\sep FCC-ee
\end{keyword}

\end{frontmatter}

\section{Introduction}
\label{sec:intro}


The Future Circular Collider (FCC-ee) will deliver enormous data samples for precision studies in Higgs, electroweak, top, and flavor physics, placing stringent requirements on detector performance to minimize experimental systematic uncertainties~\cite{FCC:2018byv, FCC:2018evy, FCC:2025lpp, FCC:2025uan, FCC:2025jtd,Dam:2025zed}. For the main tracking system, achieving the broad physics goals—such as precision Higgs measurements—requires a transverse momentum resolution of 0.1-0.2\% for charged particles at 45 GeV~\cite{FCC:2018byv}. 

To meet these requirements, the tracking detector must deliver a single-point spatial resolution of approximately 100 \micro m while maintaining a highly transparent, minimal material budget, since multiple scattering is one of the leading contributions to the momentum resolution at these momenta. 
In addition, the tracking system needs to provide excellent particle identification (PID) capability for charged pions and kaons over a wide momentum range from ${\cal{O}}(100~\text{MeV})$ to ${\cal{O}}(40~\text{GeV})$. 

Gaseous tracking detectors - such as drift chambers, straw chambers, and time projection chambers - are particularly advantageous because of their low material budget and their capability to provide both momentum measurement and PID simultaneously, without the need of a dedicated PID subsystem~\cite{FCC:2025lpp}.
In an example design of the straw tracker~\cite{FCC:2025lpp}, drift cells are constructed from cylindrical straw tubes made of 12 {\micro m}-thick metalized polyethylene terephthalate (PET) film, with diameters ranging from 1 to 1.5 cm and lengths of 4--5 meters, organized into 10 concentric superlayers. Each superlayer comprises 10 sublayers, resulting in a total of 100 layers. Each tube can measure the hit position with a spatial resolution of ${\cal{O}}$ (100 {\micro m}) for the primary coordinate in the $r-\phi$ plane. Straw tubes are arranged in both axial and stereo orientations, with stereo angles between 2$^\circ$ and 6$^\circ$, enabling hit position measurements with a precision of ${\cal{O}}(1~\text{mm})$ for the secondary coordinate along the $z$ direction. The tracker provides radial coverage from about 0.3 m to 2 meters. A Helium:isobutane (90:10) gas mixture is assumed. To enhance PID performance, a cluster-counting algorithm~\cite{Rolandi:2008ujz, Cataldi:1996mz, Elmetenawee:2025ioq} is expected to be implemented in the front-end electronics to count primary ionization clusters produced by charged particles using the full waveform information. The total material budget amounts to 1.3\% $X_0$ at $\theta=90^{\circ}$, predominantly due to the PET wall. With this detector configuration, the contribution of multiple scattering to the momentum resolution is found to be approximately 0.1\%. 

This article reports on studies of a straw tracker made of tubes with a wall thickness of 36 \micro m using 150 GeV muon beams at CERN. Since the tube wall is thicker than that proposed for the FCC-ee experiment, and because high-energy muons were used, our studies focus on the spatial resolution in both the primary and secondary coordinates, as well as on the single-tube detection efficiency.  Two distinct experimental setups were used, and their results were cross-validated for consistency. 

Existing straw-tube tracking detectors have demonstrated single-hit spatial resolutions typically in the range of $100-300$ \micro m, depending on straw diameter, gas mixture, electronics, operating conditions, and reconstruction method. Examples include the ATLAS TRT~\cite{ATLAS:2017jwu} and NA62 straw spectrometer~\cite{NA62:2017rwk}, which achieve resolutions of approximately 130 \micro m, as well as the LHCb Outer Tracker~\cite{LHCbOuterTracker2014}, COMPASS~\cite{Bychkov:2006xr}, HERMES~\cite{Ackerstaff1998HERMES}, GlueX~\cite{Adhikari2021GlueX}, PANDA~\cite{Erni2013PANDASTT}, Mu2e~\cite{Pezzullo2017Mu2eTracker}, and COMET~\cite{Nishiguchi2017COMETStraw} straw-based tracking systems, which report or target resolutions generally between 100 and 250 \micro m. 

\section{Experimental setup}
\label{sec:setup}
\begin{figure*}[ht]
    \centering
    \subfloat[]{\includegraphics[width=0.25\textwidth]{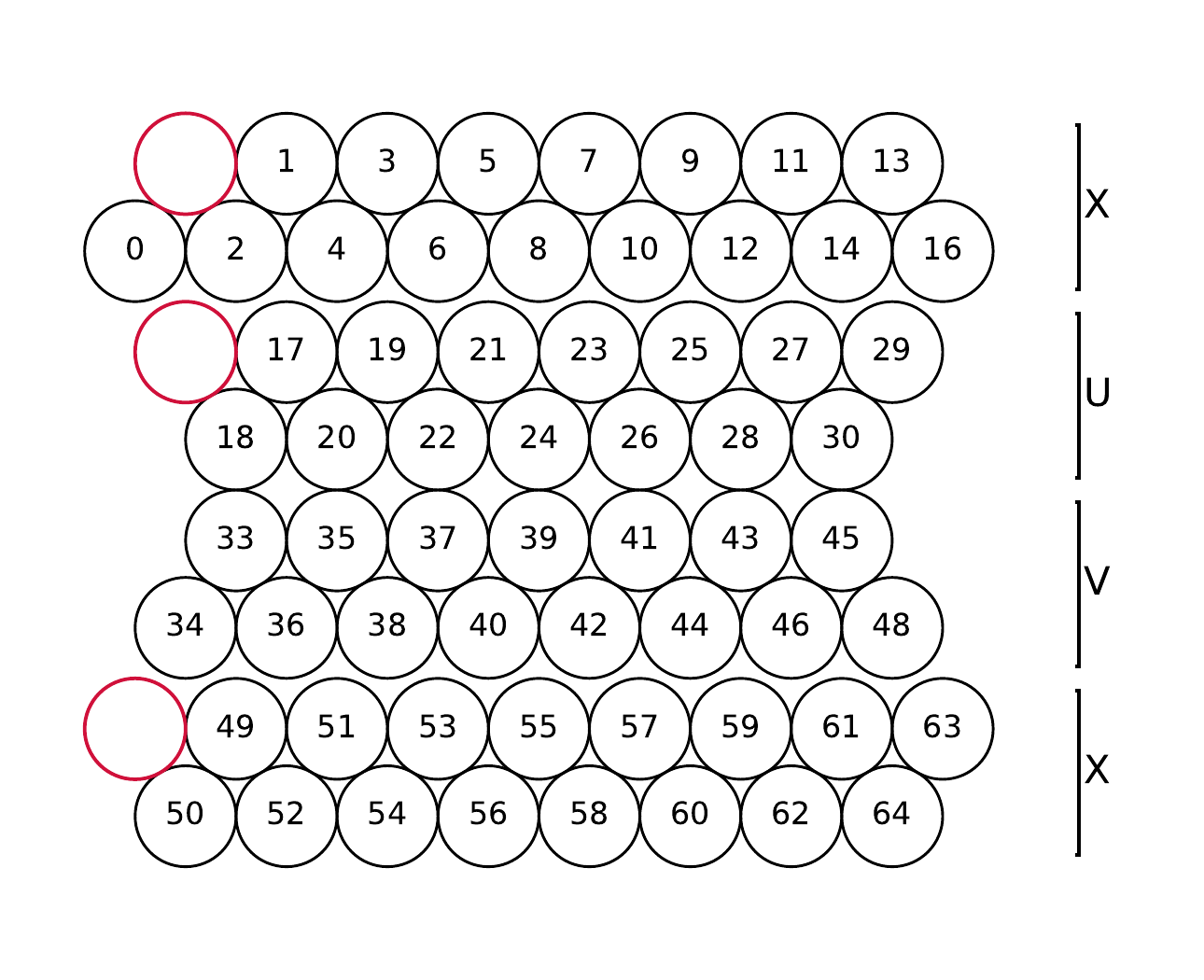} \label{fig:chamber} } 
    \subfloat[]{\includegraphics[width=0.35\textwidth]{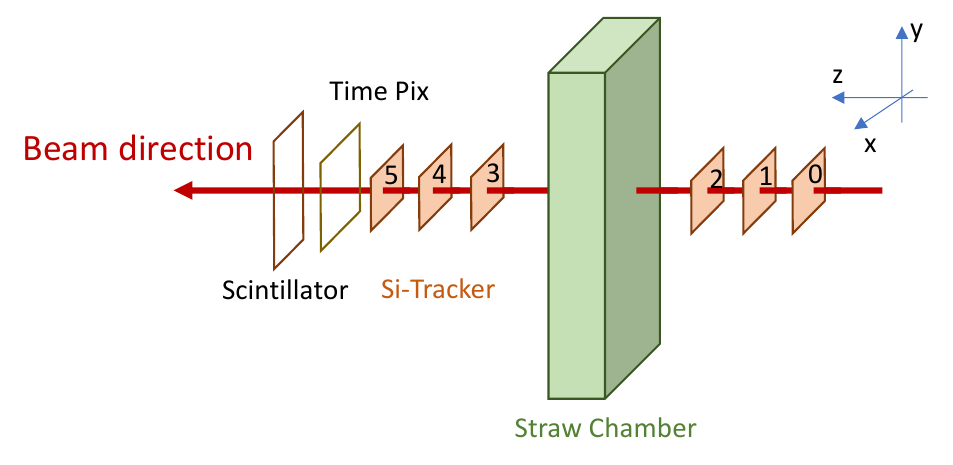} \label{fig:setup2024} } 
    \subfloat[]{\includegraphics[width=0.35\textwidth]{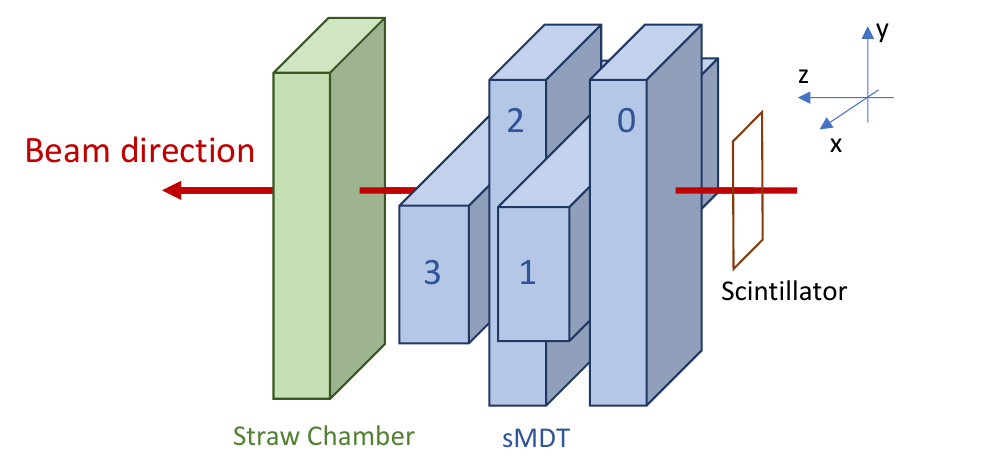} \label{fig:setup2025} } 
    \caption{(a) Layout of the straw tubes in the $x-z$ plane; (b) Layout of the experimental setup for the 2024 test beam study; and (c) Layout of the experimental setup for the 2025 test beam study.}
    \label{fig:experiment_setup}
\end{figure*}

The first test beam study was conducted in September and October 2024, followed by a second study in July and August 2025. Both experiments used a 150 GeV muon beam at CERN's H4 beam line. At this momentum, the effect of multiple scattering on the particle trajectory is negligible, allowing for a precise evaluation of the tracking detector spatial resolution without the need to account for significant scattering effects. A straw chamber produced at the Joint Institute for Nuclear Research (JINR) was used, with each straw tube featuring a diameter of $9.78 \pm 0.005$ mm and a length of 40 cm. Straw tubes were manufactured using ultrasonic welding technology that provides a sufficiently large elastic deformation range, allowing them to be stretched without damage. The tube wall thickness is 36 {\micro m} with 50 nm copper and 20 nm gold metalization on the inner surface. The chamber consists of 64 tubes arranged in 8 layers: the top two and bottom two layers contain tubes aligned along the $y$ axis, while the third and fourth layers (U-layers) are rotated by $+2^\circ$, and the fifth and sixth layers (V-layers) are rotated by $-2^\circ$. The straw tubes are glued only at the flanges and kept under slight tension.  Anode wires have a diameter of 30 {\micro m} and are tensioned to 90 grams. Wire centering is achieved through a precise fit between the pin and the corresponding hole inside the pin. 
During the beam tests, the chamber was operated with an Ar:CO$_2$ (70:30) gas mixture at 1 bar, and a high voltage of 1,750 V was applied. The layout of the straw chamber is shown in Fig.~\ref{fig:chamber}.

The tube material has the same characteristics as those used in the NA62 experiment~\cite{NA62:2017rwk}. However, several components and production techniques - including the welding technology~\cite{StrawTrackerRDteam:2025qsj}, pins, end plugs, and high-voltage interface - were significantly upgraded relative to the original NA62 production, which took place approximately 10 years earlier. The chamber’s $\pm 2^\circ$ stereo angle was motivated by the design requirements of the Spin Physics Detector experiment~\cite{Bautin:2025tdm, SPD:2024gkq}.

The experimental setup for the 2024 study is shown in Fig.~\ref{fig:setup2024}. An EUDET-type pixel beam telescope \cite{Jansen:2016}, AZALEA, comprising six high-resolution silicon sensor planes, was used as the reference tracking system, with the straw chamber positioned in the middle.
A scintillator and an additional FEI4 pixel detector were placed following the last silicon plane. The FEI4 detector~\cite{Garcia-Sciveres:2011ntv} provides triggers for the whole system with a precision of 25 ns. The scintillator, with a time resolution of $\sim 500$ ps, was introduced to provide a precise time reference signal. The AZALEA telescope provided precise reference tracks for incoming muons with a spatial resolution of about 10 {\micro m}. However, the AZALEA telescope has an active area of $10 ~\text{mm} \times 20~\text{mm}$, resulting in a reduced number of straw tubes within the acceptance of the reference tracker.

The 2025 experimental setup was similar to that used in 2024, except that the AZALEA telescope was not used. Instead, four small-diameter Monitored Drift Tube (sMDT) detectors~\cite{Amidei:2022jlx} were used to measure incoming muons, as shown in Fig. \ref{fig:setup2025}, enabling the study of more straw tubes hit by beam muons. Each sMDT chamber has 24 round aluminum tubes arranged in four layers. Each tube is 1.5 cm in diameter and 40 cm in length. The Ar:CO$_2$ (70:30) gas mixture was used at a pressure of 1 bar, and the high voltage applied was 2,100 V, yielding a spatial resolution of about 100 {\micro m} per tube. The four chambers were positioned in the $x-y$ plane, with each chamber oriented orthogonally relative to its neighbors. 

The sMDT chambers and straw tubes share a single readout system, while the AZALEA telescope was read out by a separate system. Careful synchronization between these systems was performed to ensure reliable integration of event information from all detectors.

For both sMDT and straw tubes, when a charged particle passes through a tube, ionization electrons drift toward the anode wire. The induced signal on the wire is read out by the ATLAS MDT front-end electronics.  The drift time of a muon hit is determined by the difference between the earliest arrival time of the signal at the wire and the reference time provided by the scintillator. This drift time is then used to calculate the drift radius via a Radius-Time (RT) function. Each tube's signal is first processed by an Amplifier/Shaper/Discriminator (ASD) ASIC~\cite{Kroha:2016fid, DeMatteis:2017xky}, and the timing information is digitized by a time-to-digital converter (TDC) ASIC with a bin size of 0.78 ns~\cite{Wang:2017jnd, Liang:2019weg}. The ASD ASIC is programmable with a dead time of $\sim$ 1 {\micro s} to avoid multiple hits from multiple threshold crossings of a single signal. Once the initial arrival signal is detected, no further time measurements are recorded within this dead time window.  In addition, ASD measures the signal's pulse amplitude for the first $\sim 20$ ns, enabling both gas gain monitoring and pulse-amplitude-dependent slewing corrections to the timing measurement. This amplitude is measured with a fixed rundown current on the charge of the ASD's Wilkinson integrator capacitor and is encoded as the time interval between the leading and trailing edges of the ASD output pulse. This amplitude is reported as an Analog-to-Digital Converter (ADC) value~\cite{Arai:2008zzb, Penski:2024jin}. Since the front-end electronics used here record only the initial arrival signal and then have a dead time of $\sim$ 1 \micro s, new front-end electronics must be developed for the FCC-ee straw tracker to exploit the full waveform for cluster finding. 

Each ASD ASIC processes signals from eight tubes, and each TDC handles discriminated signals from three ASDs. A mezzanine card with three ASDs and one TDC thus reads out 24 tubes in total. A Chamber Service Module (CSM) multiplexes data from up to 18 mezzanine cards and transmits it to a miniDAQ board~\cite{Guo:2022xye}, where the relevant hits are extracted from the raw data stream.
 
\begin{figure*}[ht]
    \centering
    \subfloat[]{\includegraphics[width=0.45\textwidth]{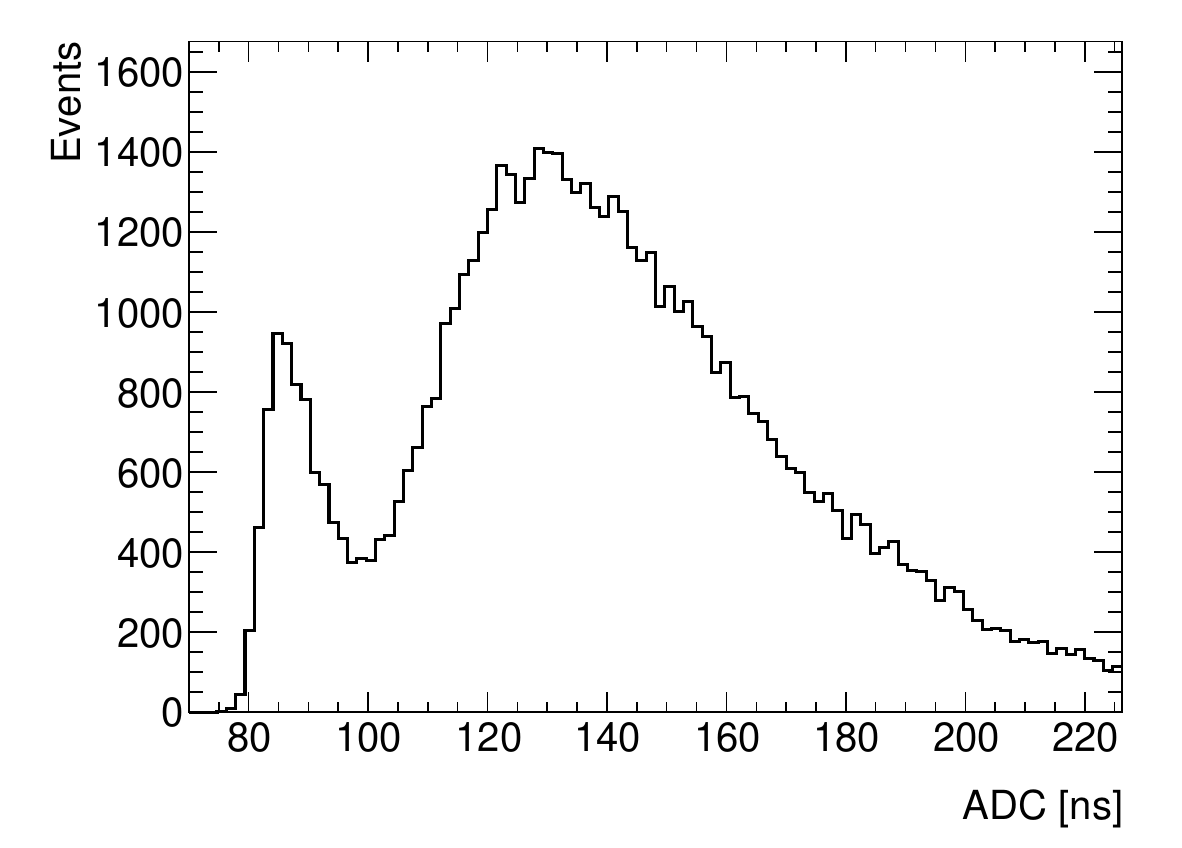} \label{fig:adc_hist}} 
    \subfloat[]{\includegraphics[width=0.45\textwidth]{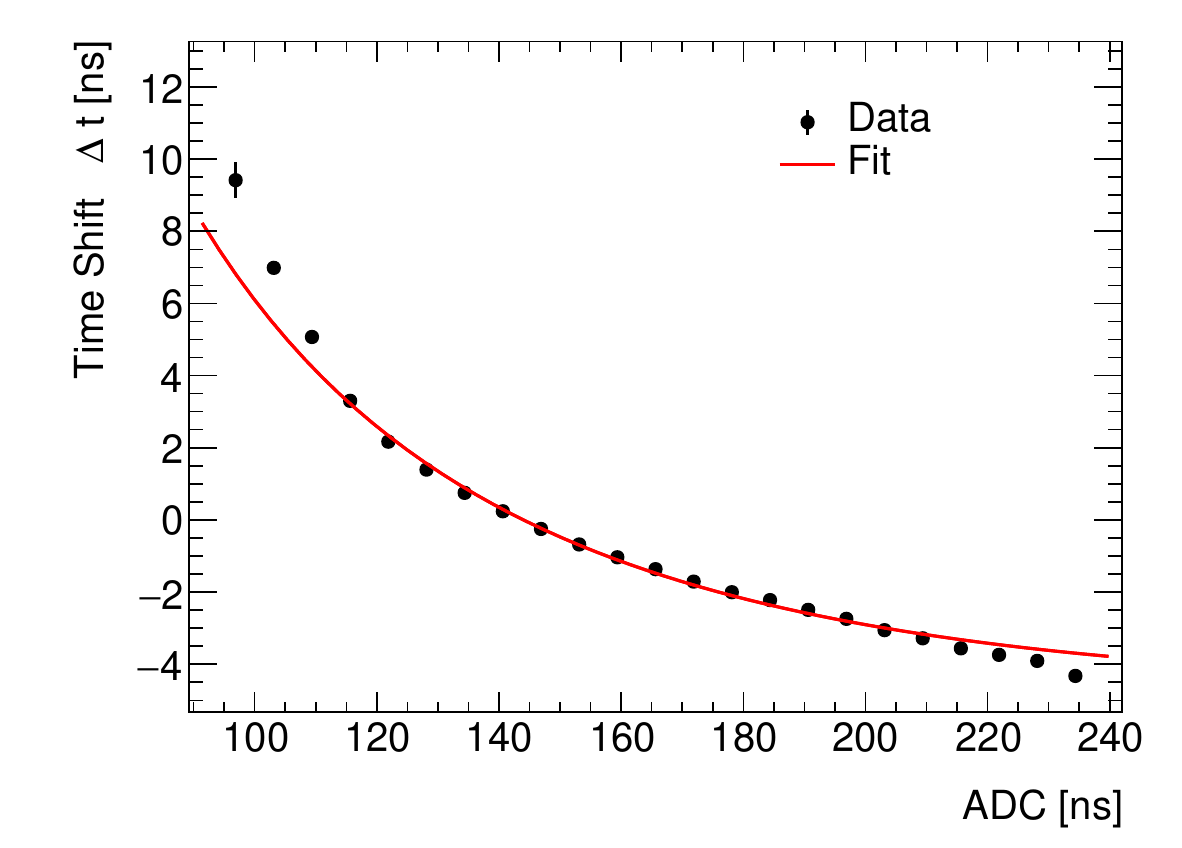}
        \label{fig:slew correction 2024}}
    \caption{(a) Raw ADC distribution for a typical channel and an ADC cut of \SI{100}{ns} was applied for this channel; and (b) drift time correction as a function of ADC value.}
\end{figure*}

Prior to the performance study of the straw chamber, a series of data pre-processing steps was applied to the raw hits. First, an optimized ADC threshold was applied to suppress noise hits. As shown in Fig.~\ref{fig:adc_hist}, the raw ADC distribution exhibits a prominent low-amplitude peak corresponding to background noise. For the data analysis, a specific cut was applied to each channel to effectively eliminate these noise hits while preserving a high acceptance for true signal hits.

Following hit selection, a time-slew correction is applied to the drift time measurements. The threshold-crossing time of the discriminator circuit depends on the amplitude of the analog signal; signals with smaller charge deposition - indicated by lower ADC values - suffer from a systematic delay known as time walk. To correct for this amplitude-dependent effect, the time delay is parameterized using the following empirical function:
\begin{equation}
    \Delta t_{\text{shift}} = p_0 + \frac{p_1}{\sqrt{\text{ADC}}}+\frac{p_2}{\text{ADC}},
\end{equation}
where $p_i$ with $i=0,1,2$ are free parameters determined by a fit to the experimental data. Figure~\ref{fig:slew correction 2024} shows the measured dependence of the drift-time correction on the ADC value. The corrected drift times were subsequently used to reconstruct drift radii and calculate the unbiased residuals.

\section{Results from the 2024 run}
\label{sec:results_2024}

\subsection{Measurements of the wire positions and angles of straw tubes}
\label{wire position}

The manufacturing process for the straw chamber introduced constraints that can lead to deviations from the ideal tube positions, particularly regarding the spacing between anode wires. Tubes in the U and V layers are rotated slightly compared to those in the X layer, and determining the actual rotation angles is important. In addition, the relative distances between the AZALEA telescope, the sMDT chambers, and the straw chamber must be measured to enable accurate extrapolation of muon tracks from the AZALEA telescope (or the sMDT chambers) to the straw array. 
Test beam data are essential for precisely determining the relative positions of all detectors.

\begin{figure*}[ht]
    \centering
    \subfloat[]{\includegraphics[width=0.47\textwidth]{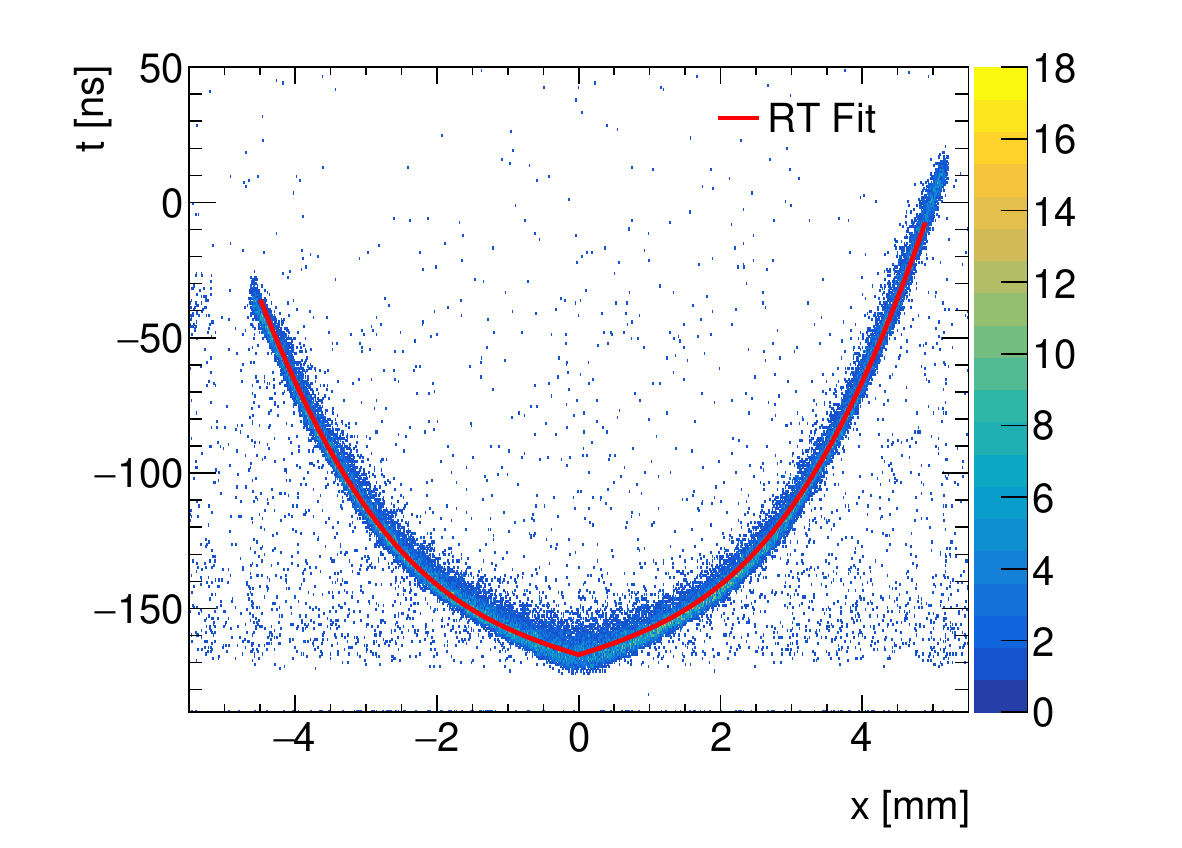} \label{fig:RT}} 
    \subfloat[]{\includegraphics[width=0.45\textwidth]{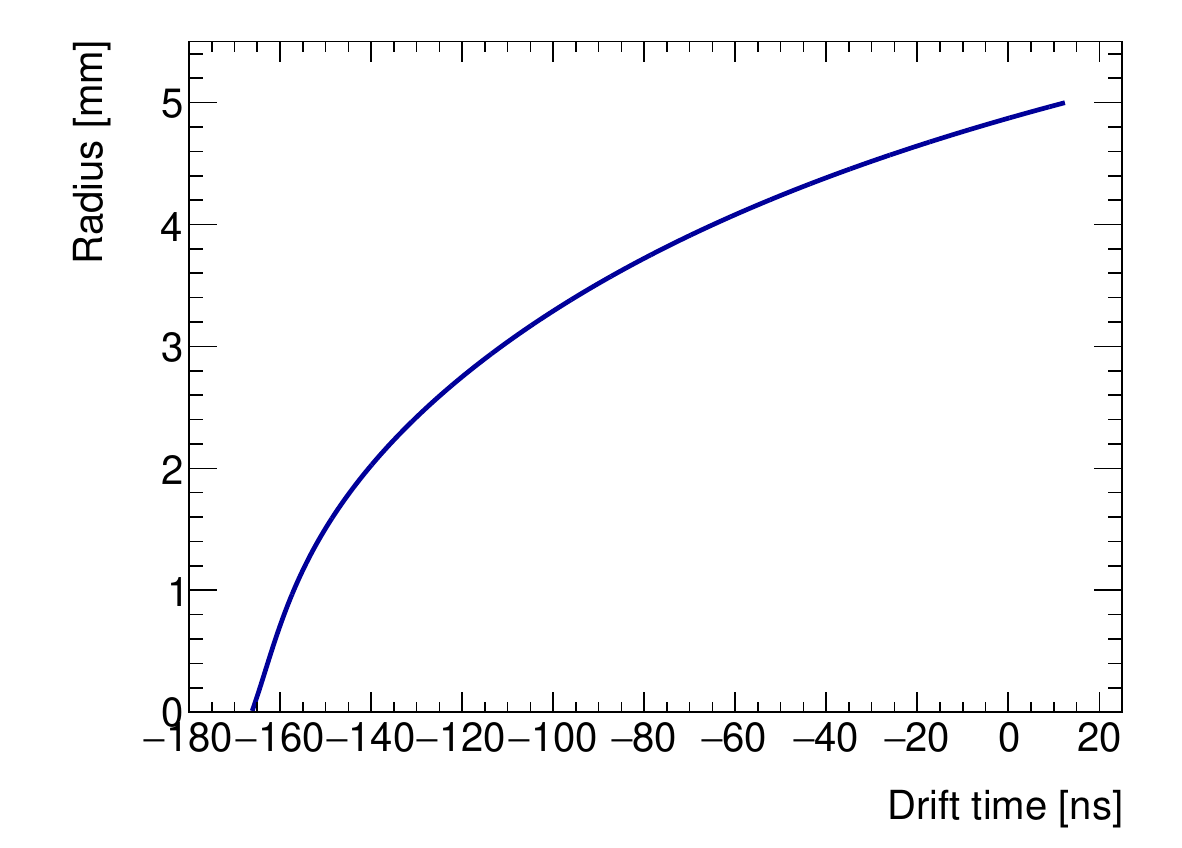} \label{fig:RT_func}} 
    \caption{(a) Drift time as a function of the extrapolated $x$ position with respect to the expected wire position, overlaid with the fitted curve using Eqn.~\ref{eq:u_shape_fit}. The observed asymmetry is attributed to the offset of the anode wire from the physical center of the tube; and (b) Average RT relation.}
    \label{fig:wire_position}
\end{figure*}

For each hit with coordinates ($x_h$, $y_h$, $z_h$), a straight line trajectory is described by $x_h=k_x z_h+b_x$ and $y_h=k_y z_h+b_y$, where $k_x$ ($k_y$) is the slope and $b_x$ ($b_y$) is the offset along the $x$ ($y$) axis. The expected $x_h$ and $y_h$ positions of a hit can be obtained by extrapolating tracks reconstructed by the AZALEA telescope to the location of the actual tube. Figure~\ref{fig:RT} shows the drift time as a function of the difference between the extrapolated position and the wire position for all hits in a single tube. This distribution exhibits a clear U-shaped profile, which is fitted using a 4${^\text{th}}$-order polynomial function centered at the wire position:
\begin{equation}
    f(x) = p_0 + p_1|x - x_0| + p_2|x - x_0|^2 + p_3|x - x_0|^3 + p_4|x - x_0|^4,
    \label{eq:u_shape_fit}
\end{equation}
where $p_i$ are the coefficients of the polynomial, and $x_0$ denotes the center of the distribution. The position of the anode wire is then determined by the fitted $x_0$ value, corresponding to the coordinate with the shortest drift time. Negative drift times occur due to the $t_0$ calibration, where the reference time provided by the scintillator system is subtracted from the raw measurements.

After determining the wire position for each tube, the drift distance is calculated as the difference between the extrapolated track position from the AZALEA telescope and $x_0$ obtained from the above fit. This drift distance, plotted against the measured drift time, is used to derive the RT function. The average RT function is shown in Fig.~\ref{fig:RT_func}. To accurately model the non-linear shape of the RT function, the curve was fitted using a spline function.

\begin{figure*}[ht]
    \centering
    \subfloat[]{\includegraphics[width=0.32\textwidth]{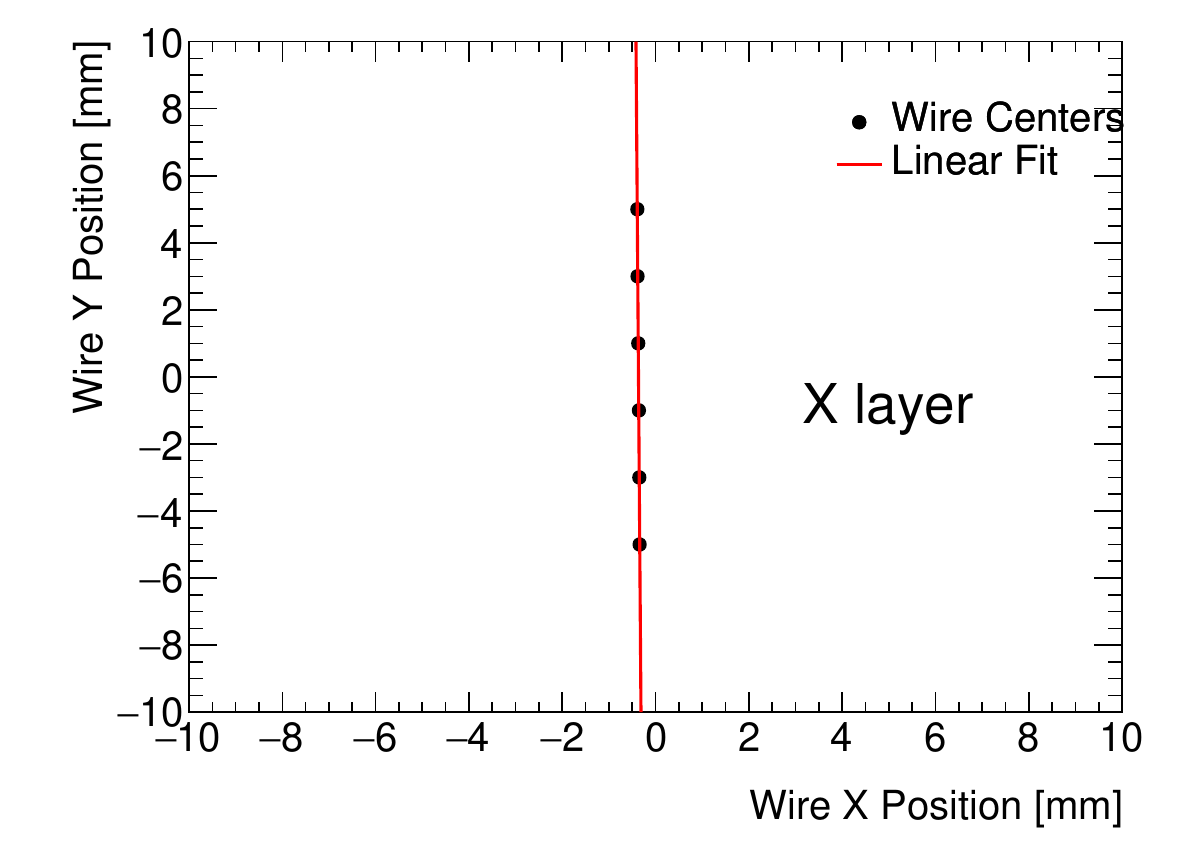} \label{fig:slope_X}} 
    \subfloat[]{\includegraphics[width=0.32\textwidth]{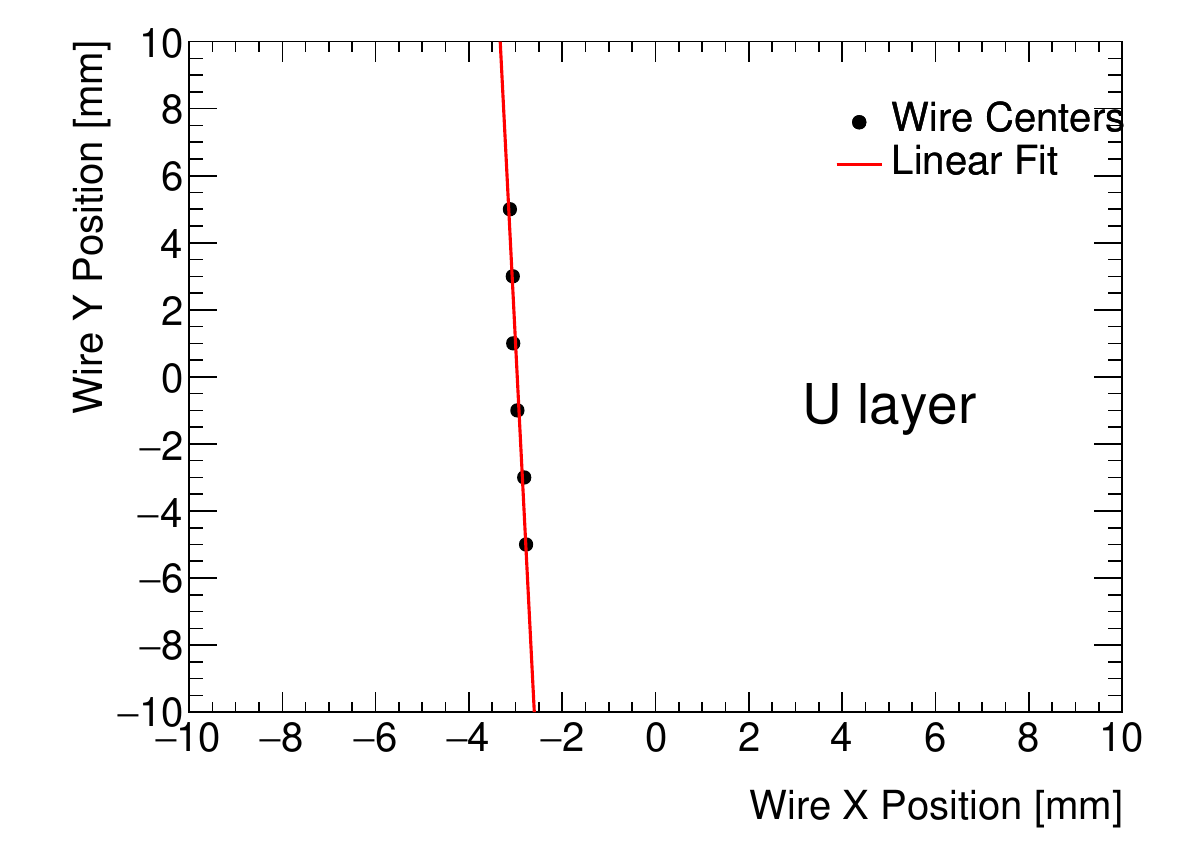}\label{fig:slope_U}} 
    \subfloat[]{\includegraphics[width=0.32\textwidth]{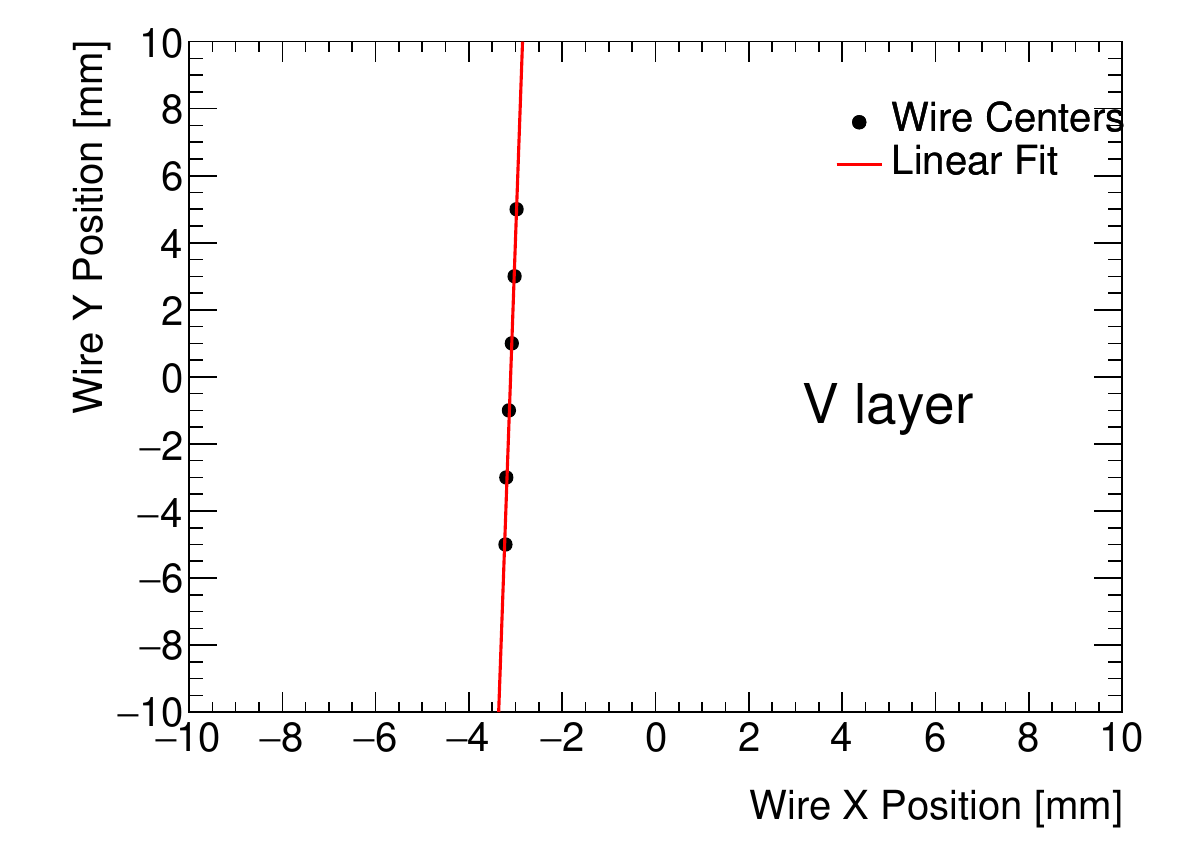} \label{fig:slope_V}} 
    \caption{Dependencies of the extracted wire position as a function of the track $y$ coordinate for the (a) X-, (b) U-, and (c) V-layers. The data points from each $y$ coordinate subset are modeled using a linear fit to determine the tube inclination angles.}
    \label{fig:tube_angle}
\end{figure*}

To determine the rotation angles between the straw tubes and the $y$-axis, the analysis was performed for tracks at different $y$-coordinate values separately, and the wire position was found from the RT curves of each data subset. Figures~\ref{fig:slope_X}, \ref{fig:slope_U} and \ref{fig:slope_V} show the dependencies of the wire position as a function of the track $y$ coordinate for X-, U-, and V-layers, respectively.
The fitted tube angles for the X-, U-, and V-layers were found to be  0.4$^\circ$, -1.49$^\circ$, and 2.49$^\circ$, respectively. The values indicate that the straw chamber has a general inclination of about 0.4$^\circ$ with respect to the reference tracker.

\subsection{Single tube spatial resolution}
To estimate the spatial resolution for a straw tube, an unbiased hit residual ($\sigma_{\text{unbiased}}$) is calculated. The unbiased residual is defined as the difference between the extrapolated track position - determined solely from the AZALEA telescope - and the measured drift radius from the straw tube. The hit prediction resolution ($\sigma_{\text{hit}}$) represents the uncertainty of the extrapolated track position, which depends on the spatial resolution of the AZALEA telescope and the extrapolation distance. The spatial resolution of the straw tube is then determined as follows:

\begin{equation}   
\sigma = \sqrt{\sigma_{\text{unbiased}}^2 -\sigma_{\text{hit}}^2}.
\label{eq: resolution}
\end{equation} 

\begin{figure*}[ht]
    \centering
    \subfloat[]{\includegraphics[width=0.45\textwidth]{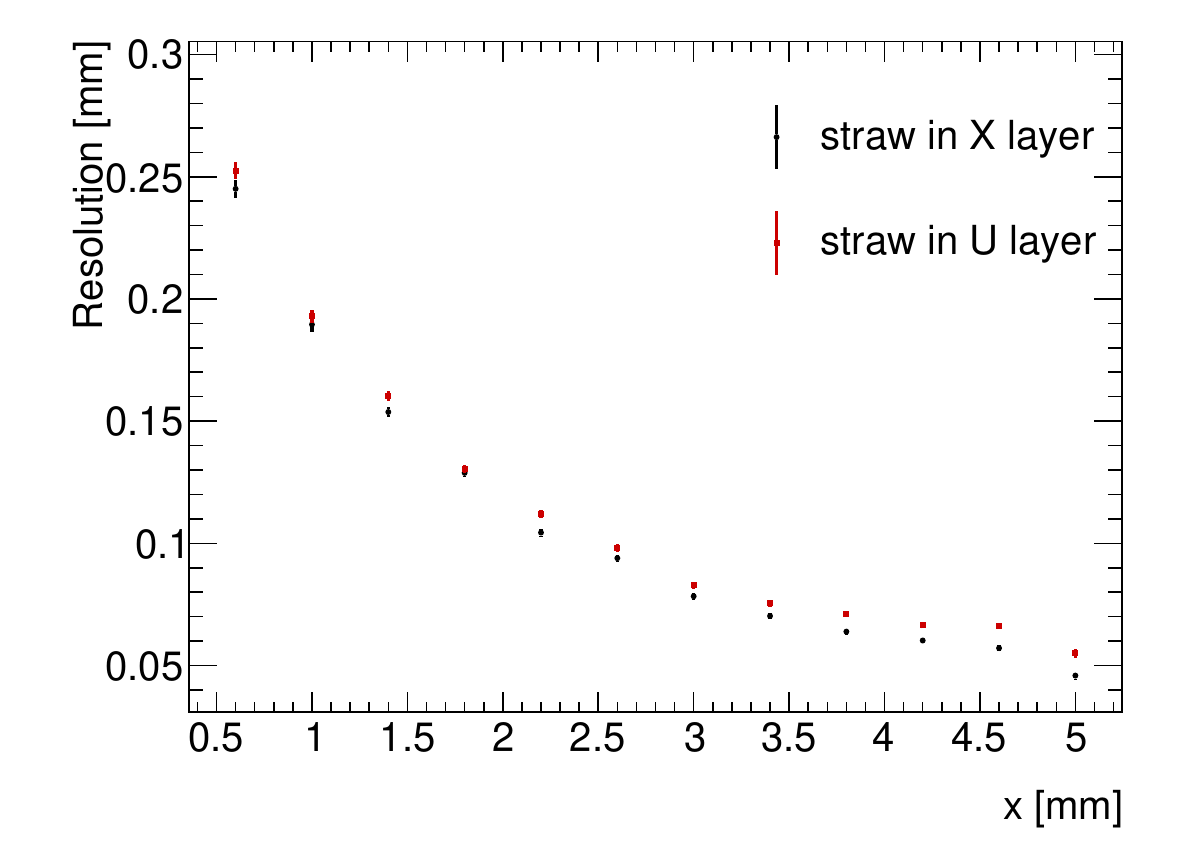}
    \label{fig:resolution_primary}}
    \subfloat[]{\includegraphics[width=0.45\textwidth]{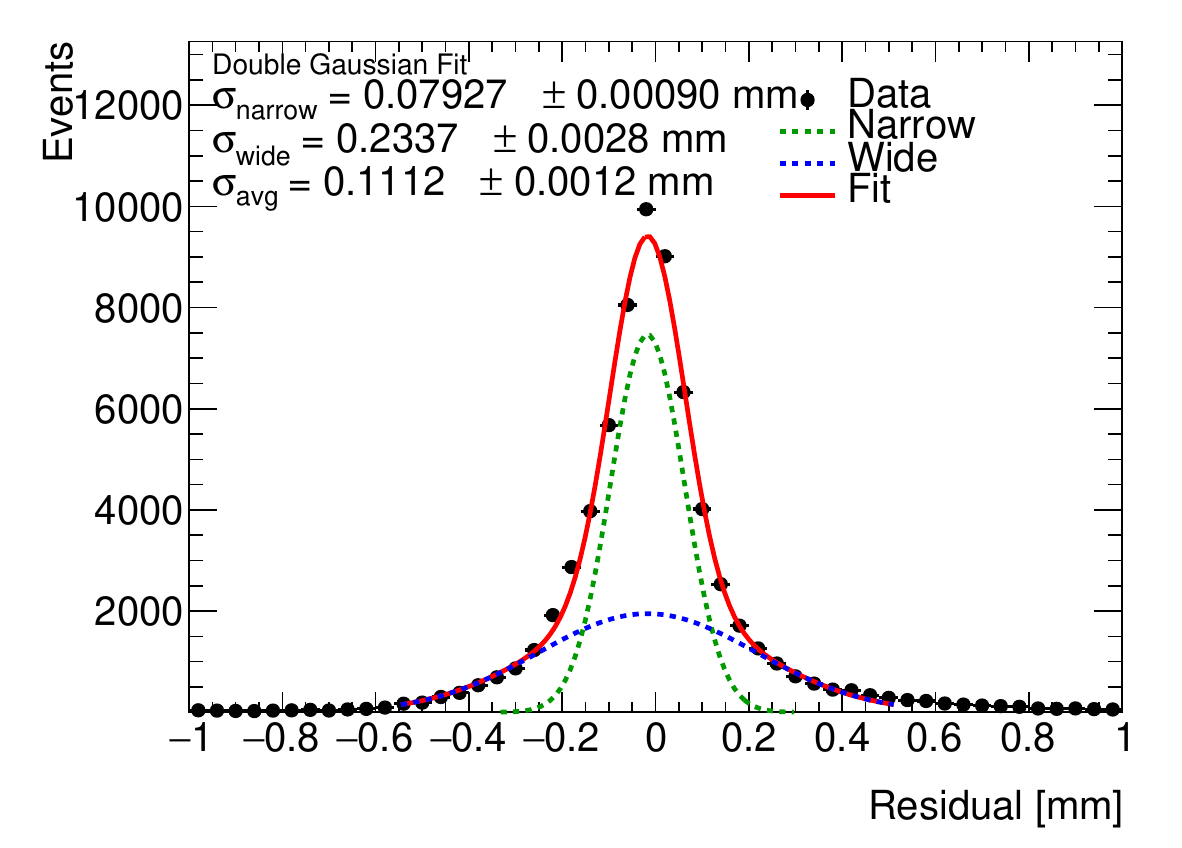}
    \label{fig:average resolution}}
      \caption{(a) Single tube spatial resolution as a function of the drift radius for two typical straw tubes; and (b) The residual distribution for all tubes under study. A double Gaussian function was used to fit the data to obtain the average spatial resolution.}
    \label{fig:resolution}
\end{figure*}

The AZALEA telescope has an excellent spatial resolution of better than 10 \micro m, while the $t_0$ reference uncertainty is around 500 ps. Since these intrinsic uncertainties are significantly smaller than $\sigma_{\text{unbiased}}$, and the straw chambers were centrally located within the AZALEA telescope, the track prediction uncertainty is practically negligible. Figure~\ref{fig:resolution_primary} shows the spatial resolution as a function of the drift radius for two typical straw tubes. The resolution improves near the edge of the tube, where the longer drift time allows for more precise time measurements, but deteriorates when the muon passes near the anode wire. Figure~\ref{fig:average resolution} shows the residual distribution for all tubes under study. The data are fitted with a double Gaussian function. The average width, $\sigma_{\text{avg}}$, is calculated as the area-weighted mean of the narrower component ($\sigma_{\text{narrow}}$) and the wider component ($\sigma_{\text{wide}}$). The average spatial resolution is found to be $111.2 \pm 1.2$ {\micro m}. Our results can be compared with those of the NA62 straw tracker~\cite{Sergi:2012twa, NA62:2017rwk}, which used the same tube material and anode wire but operated at atmospheric pressure inside a vacuum tank. The NA62 straw tracker achieved a spatial resolution of approximately 200 {\micro m} on average and 130 {\micro m} near the tube wall.

\subsection{Spatial resolution along the tube direction}
\label{sec:2nd 2024}

The configuration of the straw chamber with four X-layers, two U-layers, and two V-layers enables reconstruction of the hit position along the tube direction, known as the secondary coordinate. A track reconstruction algorithm in three-dimensional space is needed. Only tubes aligned with the expected trajectory, as predicted by the tracks reconstructed using the AZALEA telescope, were included in the three-dimensional track reconstruction. 

\begin{figure*}[ht]
    \centering{\includegraphics[width=0.45\textwidth]{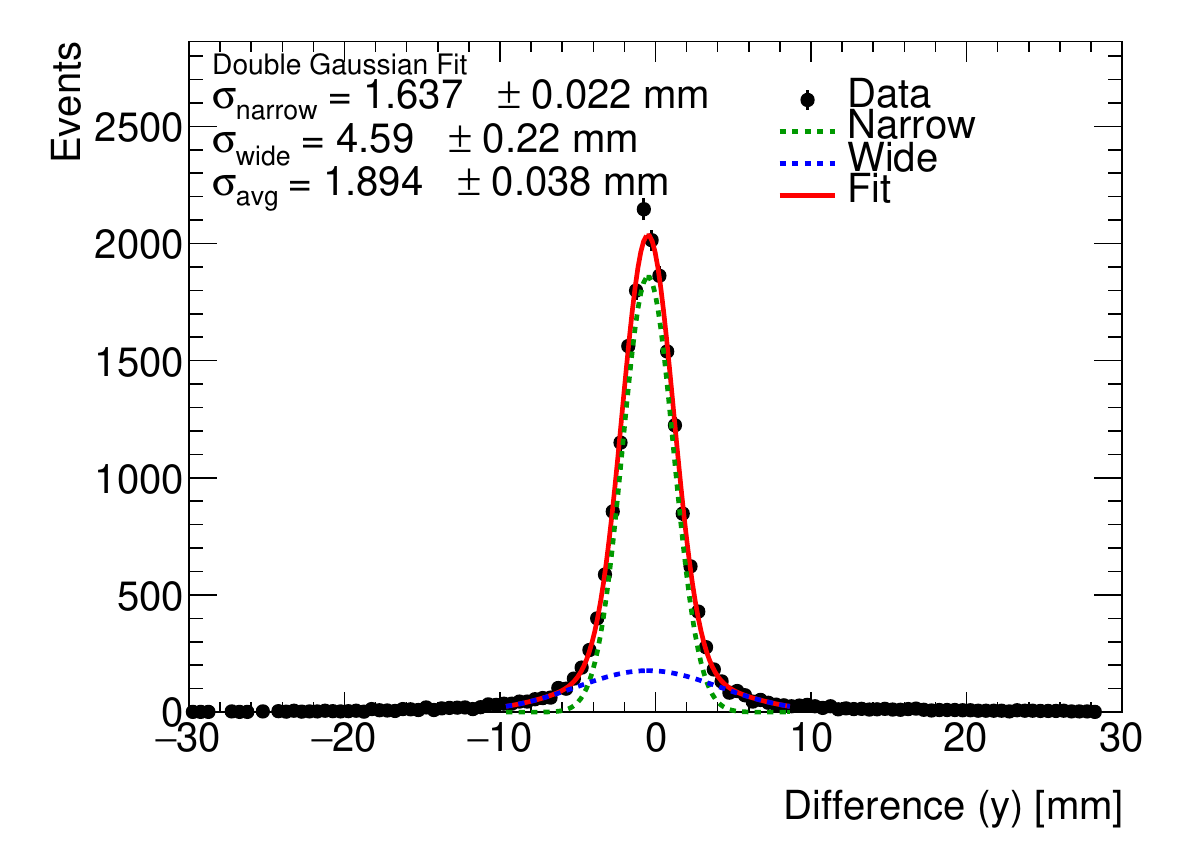} }
    \caption{The residual distribution for the secondary coordinate measurement.}
    \label{fig:2nd resolution centre}
\end{figure*}

When a charged particle passes through the chamber, its trajectory can be described as the common tangent to a set of drift circles determined by the tubes it hits. The three-dimensional distance between the hypothesized track and the anode of a straw tube is calculated as:

\begin{equation}
d_i = \left| \frac{A_i(k_yZ_i+b_y)+B_i - (k_xZ_i+b_x)}{\sqrt{A_i^2+1+(k_x-A_ik_y)^2}} \right|,
\label{eq: 3d distance}
\end{equation}
where $i = 1, \cdots, N$ with $N$ as the total number of tubes with hits, $A_i$ and $B_i$ represent the slope and intercept of the anode wire of the $i$-th tube in the plane perpendicular to the direction of the beam, and $Z_i$ denotes the position of the tube along the beam direction. 

The goal is to find a straight line with $x = k_xz+b_x$ and $y = k_yz+b_y$ that passes at a distance $R_i$ from each wire position. The optimal parameter values of $k_x$, $k_y$, $b_x$, and $b_y$ are determined by minimizing the sum of squared differences between the measured drift circles ($R_i$) and the calculated distances ($d_i$):
\begin{equation}
S (k_x,k_y,b_x,b_y) = \sum_{i=1}^N \left ( \frac{\delta_i d_i-R_i}{\sigma_i} \right)^2,
\end{equation}
where $\sigma_i$ is the uncertainty associated with each measurement (based on the results shown in Fig. \ref{fig:resolution_primary}). The parameter $\delta_i$, which takes the value of either $+1$ or $-1$, indicates whether the particle passes to the right or left side of the tube. 

Successful minimization requires appropriate initial values for $k_x$, $b_x$, $k_y$, $b_y$, and $\sigma_i$. In this beam test, the particle trajectories are well aligned with the $z$-axis, with angular divergence less than \SI{0.001}{rad}. Therefore, the possible transverse positions as the particle passes through the tube plane can be approximated by $B_i+R_i$ or $B_i-R_i$. To estimate initial values for $k_x$ and $b_x$, one can iterate over possible hit sides for each tube, perform least squares fitting, compute the corresponding $\chi^2$ values, and select the configuration with the lowest $\chi^2$ value. Using the fitted $k_x$, $b_x$, and the values of $R_i$ for tubes in the U- and V-planes, a similar procedure yields initial values for $k_y$ and $b_y$ in the $y$-direction. With these initial fitting parameters, the three-dimensional track can be reconstructed by minimizing $S(k_x,b_x,k_y,b_y)$. 

It should be noted that this $\chi^2$ minimization approach assumes a straight-line trajectory, which is effective for the 150 GeV high-momentum muons analyzed in this study. However, for lower-momentum particles where multiple Coulomb scattering becomes a significant factor, this straight-line approximation would no longer be valid. In such cases, more sophisticated track fitting algorithms, such as the Kalman filter, would be required to account for multiple scattering effects properly.

Once three-dimensional tracks are reconstructed for the straw chamber, the hit resolution along the tube direction can be evaluated. The residual is calculated as the difference between the $y$ coordinate obtained from the AZALEA telescope’s reconstructed track and that from the straw chamber. The resulting residual distribution is shown in Fig.~\ref{fig:2nd resolution centre}. A double Gaussian function was used to fit the data. Similarly, the average width is calculated as the area-weighted mean of the narrower component ($\sigma_{\text{narrow}}$) and the wider component ($\sigma_{\text{wide}}$). The average spatial resolution is found to be $1.89 \pm 0.04$ mm. 

\subsection{Single tube detection efficiency}
\label{detection efficiency}
For the efficiency determination, a tube is considered to have a valid hit if any signal crossing the optimized ADC threshold is recorded, regardless of whether the measured drift radius perfectly matches the extrapolated track distance. Since the noise rate was highly suppressed by the ADC cuts, as discussed in Section \ref{sec:setup}, the contribution from random-noise false hits is negligible.

\begin{figure*}[t]
    \centering
    \includegraphics[width=0.45\linewidth]{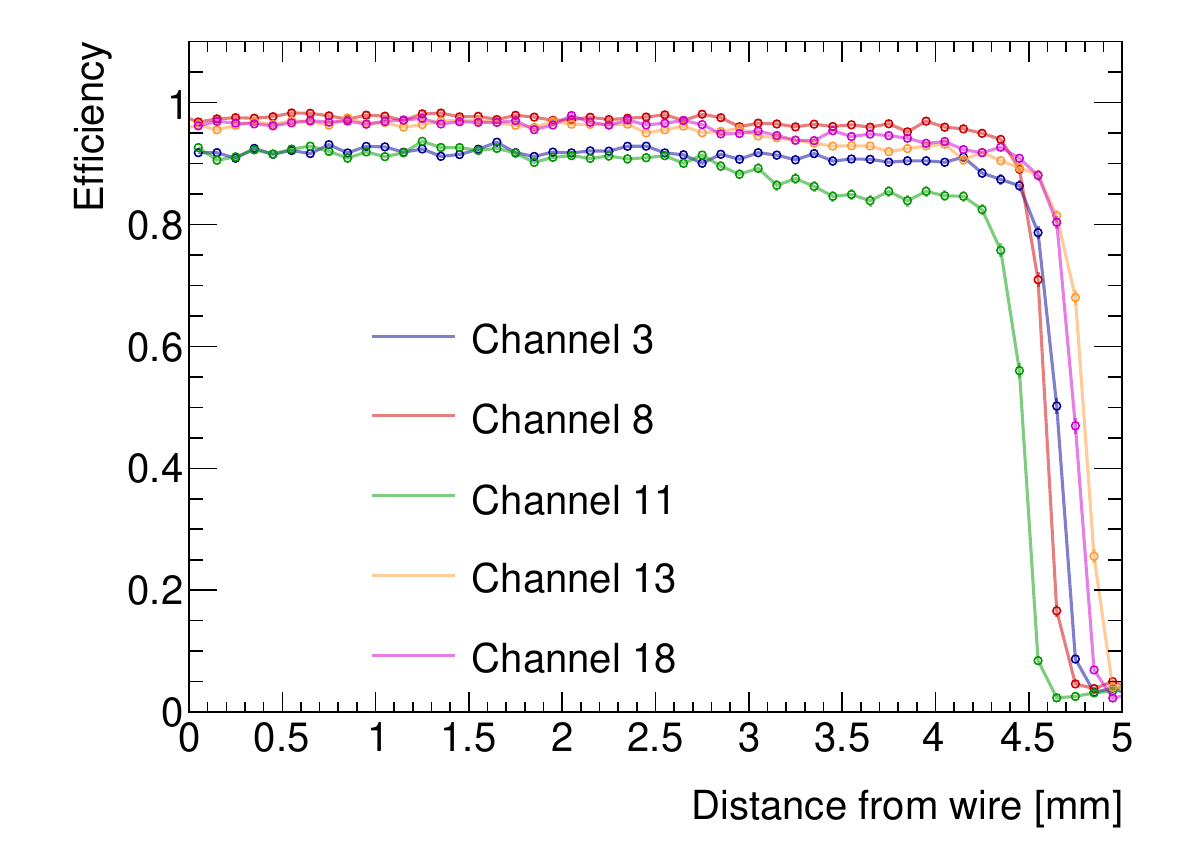}
    \caption{Detection efficiency as a function of the extrapolated position from tracks reconstructed by the AZALEA telescope for five straw tubes.}
    \label{fig:efficiency}
\end{figure*}

The single tube efficiency was also measured as a function of the drift radius. For each tube, the efficiency is defined as the ratio of the number of hits recorded for that tube and the total number of tracks passing through that tube, while the same ADC cuts were applied as mentioned in Fig.~\ref{fig:adc_hist}. The efficiency curve as a function of the extrapolated position from tracks reconstructed by the AZALEA telescope is shown in Fig.~\ref{fig:efficiency} for five straw tubes. A few tubes have slightly reduced efficiency due to insufficient gas gain. 
Due to the limited geometric acceptance of the AZALEA telescope used in the 2024 run, only 14 tubes could be evaluated for efficiency and only a subset of them were fully illuminated by the muon beam across their entire diameter. To ensure sufficient statistics and obtain reliable efficiency curves for a larger number of tubes, the hit positions were folded around the anode wire. Consequently, the efficiency is presented as a function of the absolute distance from the wire.

Six straw tubes demonstrated high efficiency ranging from 96\% to 98\%, while seven exhibited an efficiency between 90\% and 96\%. One tube, characterized by a higher noise level, yields a lower efficiency of 81\%. For reference, the NA62 straw tracker reported a single-tube efficiency of 99\%\cite{Sergi:2012twa}. 

\section{Results from the 2025 run}
\label{sec:results_2025}

In the 2025 run, sMDT chambers were used instead of AZALEA telescope. The large geometric acceptance of the sMDT chambers allows extrapolated tracks to cover 23 tubes rather than 14. However, the limited spatial resolution of the sMDT chambers poses challenges on the data analysis. 

\subsection{Single tube spatial resolution }
The spatial resolution of the sMDT chambers, previously measured in the laboratory using cosmic muons, is approximately 110 {\micro m} per tube, which is comparable to that of the straw tubes. As a result, the extrapolated position for each straw tube, based on tracks reconstructed by the sMDT chambers, has a larger uncertainty compared to positions determined using tracks from the AZALEA telescope used in the 2024 run. This extrapolation uncertainty must be excluded when estimating the spatial resolution of each straw tube. This uncertainty was evaluated using a simulation program. 

\begin{figure*}[ht]
    \centering
    \subfloat[]{\includegraphics[width=0.32\textwidth]{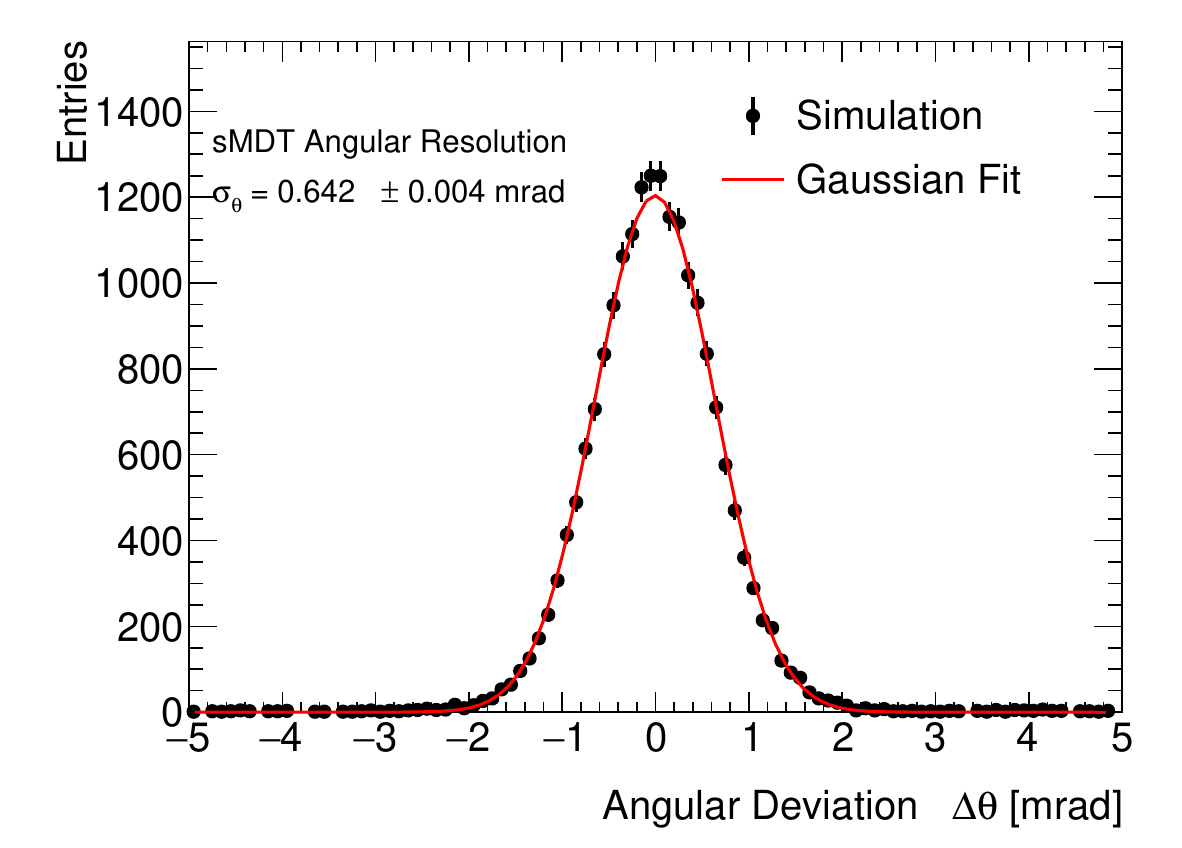} \label{fig:angular resolution}}
    \subfloat[]{\includegraphics[width=0.32\textwidth]{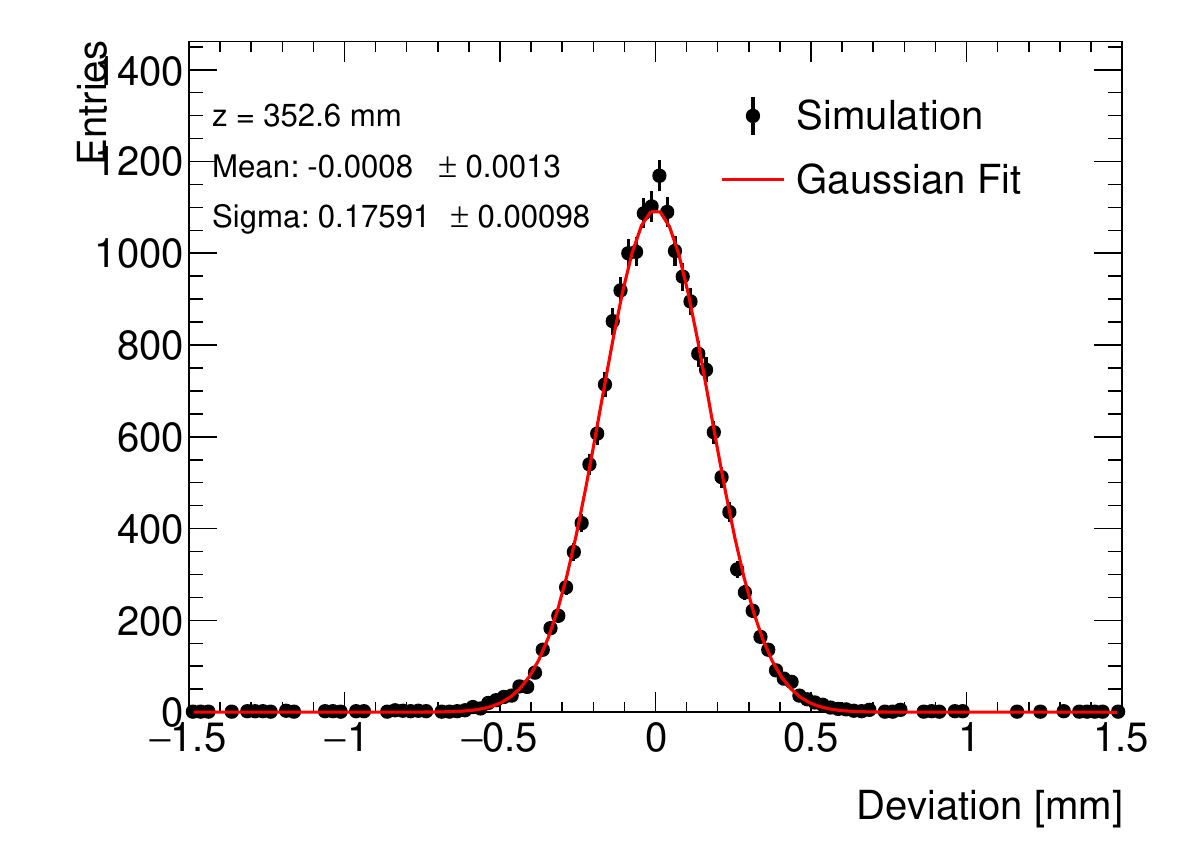} \label{fig:hit resolution}}
    \subfloat[]{\includegraphics[width=0.32\textwidth]{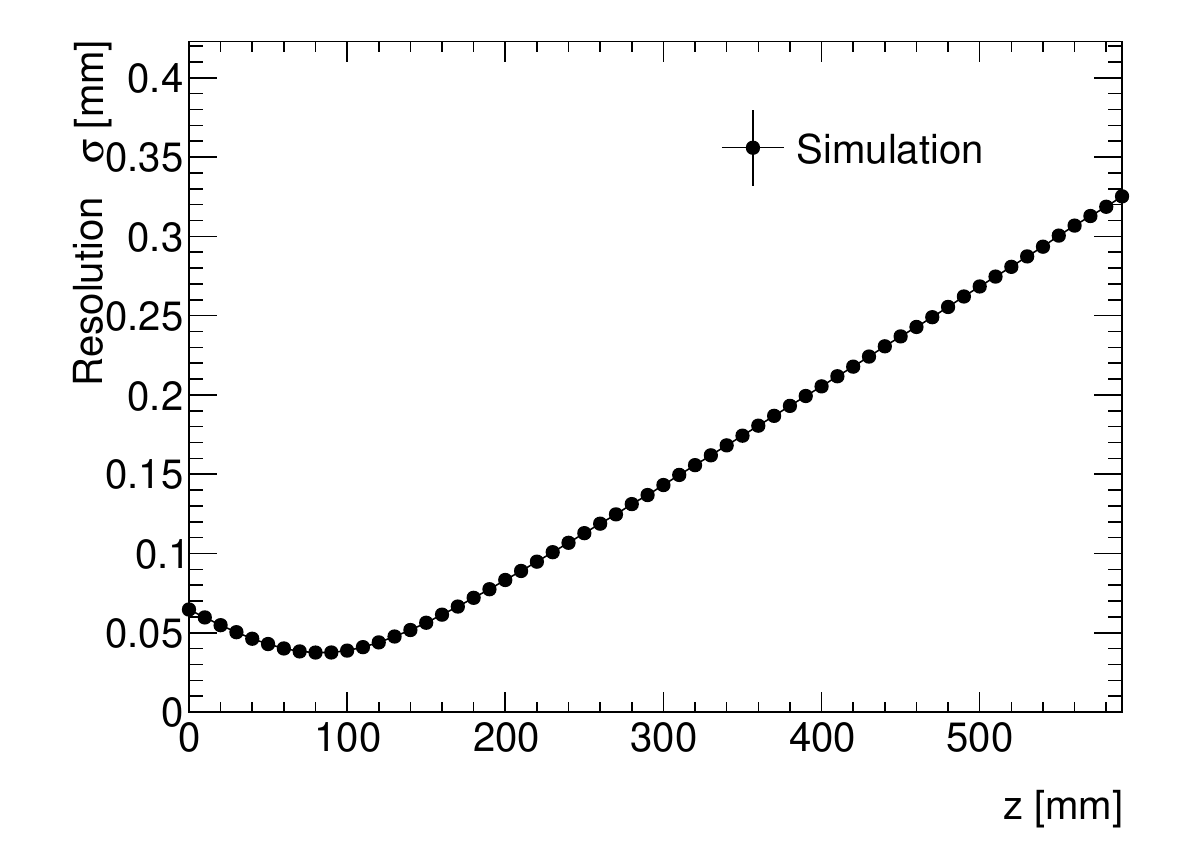} \label{fig:hit resolution vs distance}} 
    \caption{ (a) Distribution of the angular deviation; (b) Distribution of the extrapolated position deviation at the center of the straw tube detectors; and (c) Dependence of the extrapolated position resolution as a function of the $z$ position, where $z = 0$ corresponds to the rightmost surface of the sMDT chambers shown in Fig.~\ref{fig:setup2025}.}
    \label{fig:resolution vs z 2nd 2024}
\end{figure*}

In the simulation, a muon traverses several sMDT tubes, and the drift radii are smeared according to the sMDT’s measured drift-radius-dependent spatial resolution. A track fit is then performed using all the smeared sMDT drift radii. The angular deviation, defined as the difference between the true muon direction and the reconstructed track direction, is used to estimate the extrapolated position uncertainty. The angular deviation distribution is shown in Fig.~\ref{fig:angular resolution}. A fit using a Gaussian distribution is performed and the resolution is found to be approximately 0.64 mrad. This angular uncertainty is then extrapolated to straw tubes located at various $z$ positions. Defining the front surface of the first sMDT layer as the origin, the center of the straw chamber is located at $z=352.6 \pm 5.0 $ mm. The resulting average extrapolated position resolution is $175.9 \pm 3.2$ \micro m, as shown in Fig.~\ref{fig:hit resolution}. This resolution depends on the $z$ position of the straw tubes, as shown in Fig.~\ref{fig:hit resolution vs distance}.

The sMDT chambers' planes were oriented nearly perpendicular to the beam direction during the 2025 test beam runs. As a result, beam particles could hit several tube walls and wires at the same time, creating regions of inefficiency in the reference track reconstruction. 



\begin{figure*}[ht]
    \centering
    \subfloat[]{\includegraphics[width=0.45\textwidth]{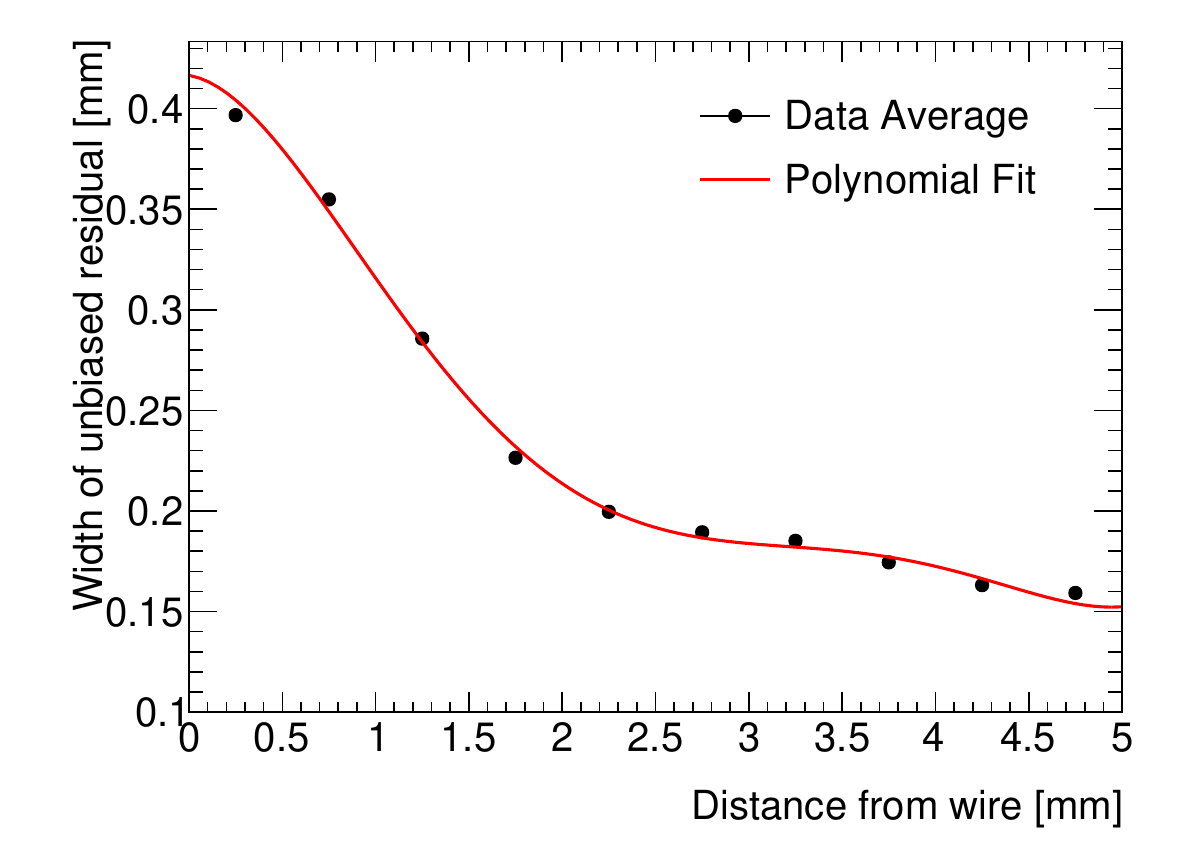} \label{fig:2025 unbiased residual vs radius}}
    \subfloat[]{\includegraphics[width=0.45\textwidth]{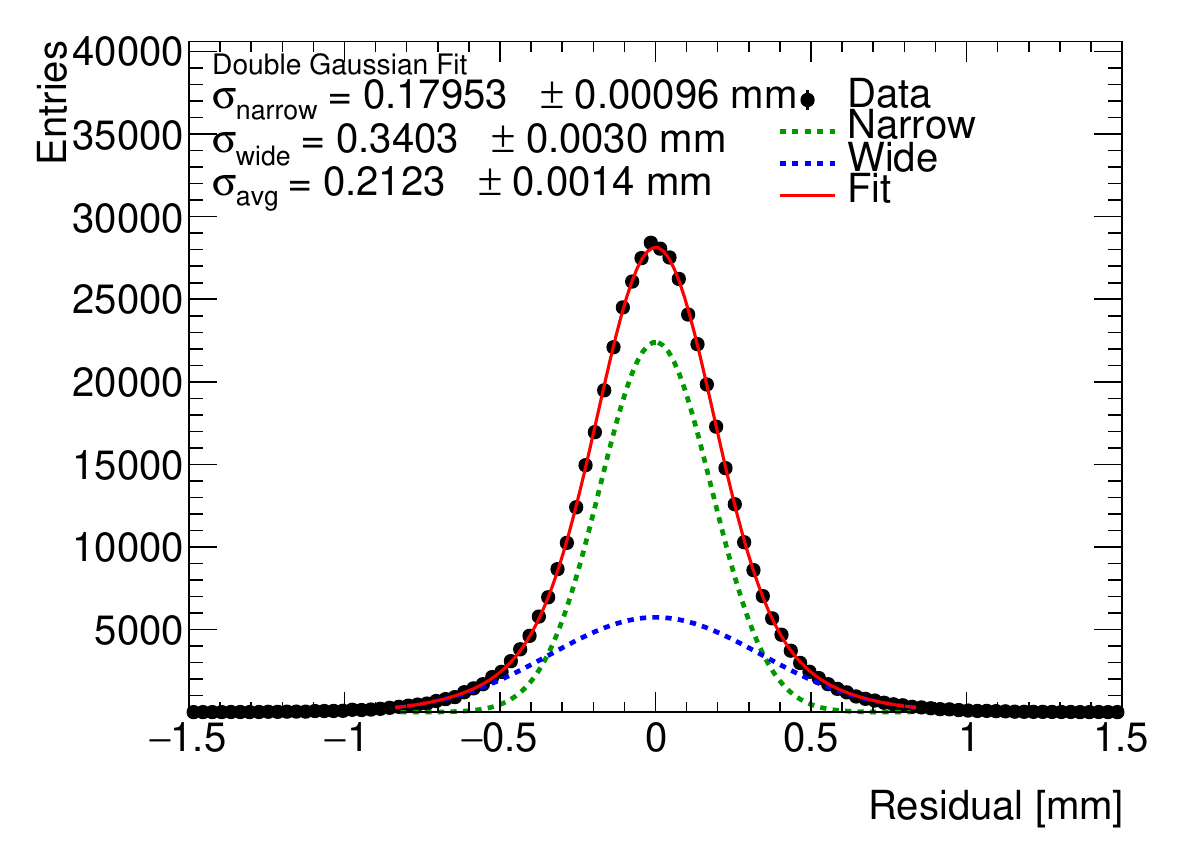} \label{fig:average unbiased residual}} 
    \caption{(a) Average width of the unbiased residual as a function of the extrapolated position; and (b) Distribution of the average unbiased residual after taking into account inefficient regions.}
    \label{fig:unbiased residual in 2025}
\end{figure*}

The unbiased residual in data is calculated as the difference between the measured straw tube hit positions and the extrapolated positions from tracks reconstructed using the sMDT chambers. Figure~\ref{fig:2025 unbiased residual vs radius} shows the averaged Gaussian-fitted widths of the unbiased residual for all 23 tubes as a function of the extrapolated position. To ensure physically reasonable behavior near the wire and tube wall, the polynomial fit to the unbiased residual width-versus-position profile was constrained to be monotonically decreasing in the boundary intervals of $0 - 0.25$ mm and $4.75 - 5.0$ mm. Since the inefficient regions distort the measurement, a simple average would not accurately reflect the actual resolution. A toy model simulation was developed to extract the average resolution. We assume a uniform hit rate across the tubes and generate hit positions using the dependence shown in Fig.~\ref{fig:2025 unbiased residual vs radius}. Inefficiency regions derived from the data are then applied. The distribution of the resulting unbiased residual is shown in Fig.~\ref{fig:average unbiased residual}. Similar to Fig.~\ref{fig:average resolution}, a double Gaussian fit is performed and the average resolution of the unbiased residual is found to be approximately 212 {\micro m}.

After subtracting an extrapolated position resolution of 175.9 {\micro m} from tracks reconstructed using the sMDT chambers, the actual average spatial resolution of the straw chamber is between 114 {\micro m} and 123 {\micro m}, consistent with the results obtained in 2024.

\subsection{Detection efficiency and tube wire offsets}
\begin{figure*}[ht]
    \centering
    \subfloat[]{\includegraphics[width=0.45\textwidth]{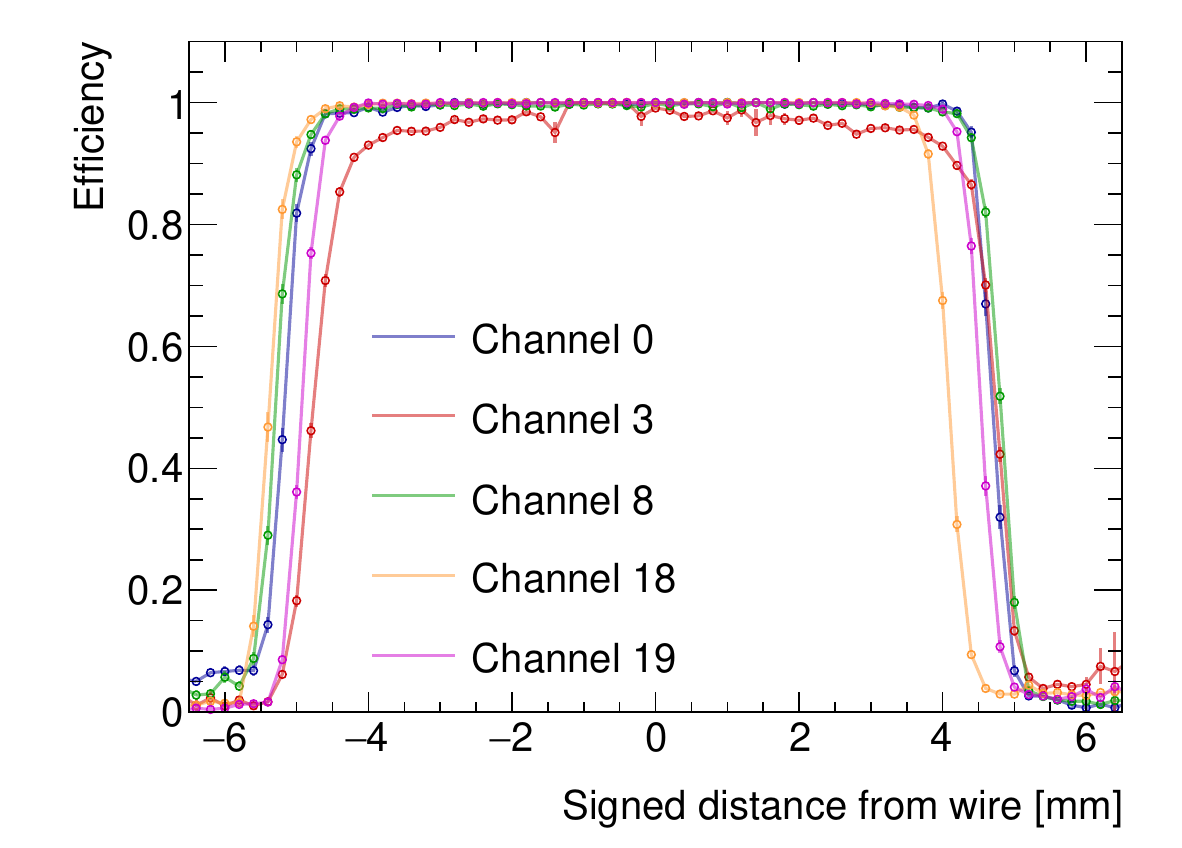} \label{fig:eff2025}}
    \subfloat[]{\includegraphics[width=0.45\textwidth]{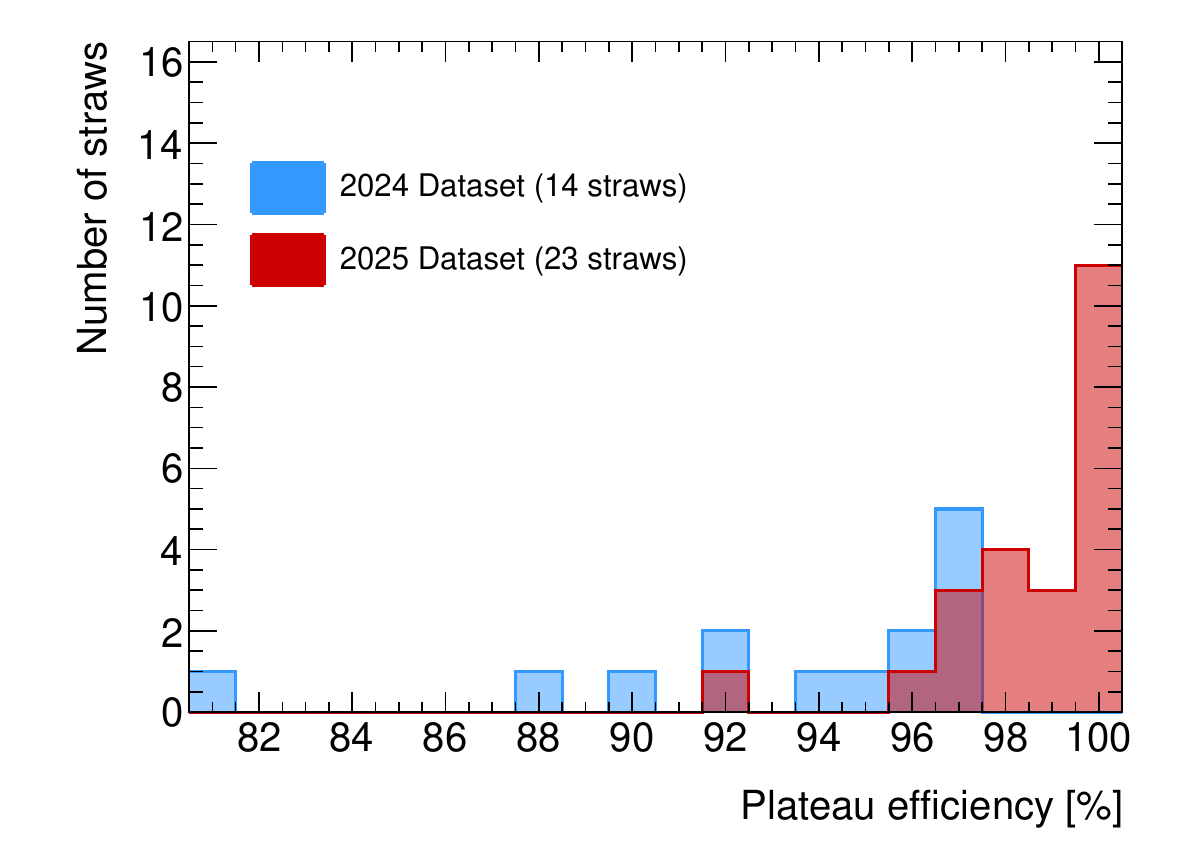} \label{fig:eff2024-2025}} 
    \caption{(a) Detection efficiency as a function of the extrapolated position, determined using tracks reconstructed by the sMDT chambers, for five straw tubes in the 2025 run; and (b) plateau efficiency for the tubes studied in the 2024 and 2025 runs.}
    \label{fig:efficiency 2025}
\end{figure*}

The larger geometric acceptance of the sMDT chambers allows us to investigate a total of 23 straw tubes in the 2025 run. The efficiency was estimated using the same procedure described in Sect.~\ref{detection efficiency}. Figure~\ref{fig:eff2025} illustrates the efficiency as a function of the hit position with respect to the location of the anode wire derived from the RT relation established in Sect.~\ref{wire position}. Of the 23 tubes evaluated, 18 exhibit high plateau efficiency above 98\%, and 4 tubes show efficiency between 96\% and 98\%. A single outlier, characterized by a higher noise level, yields a lower efficiency of 92\%. Figure~\ref{fig:eff2024-2025} compares plateau efficiency for the tubes studied in the 2024 and 2025 runs. The improvement in detection efficiency observed in 2025 is mainly due to the higher electronics threshold used, which was changed from \SI{-28}{mV} to \SI{-40}{mV}. This change suppressed the noise rate and reduced the low-ADC noise peak shown in Fig. \ref{fig:adc_hist}. As a consequence, a lower ADC cut was applied to reject noise while retaining more real events.

\begin{figure*}[ht]
    \centering
    \subfloat[]{\includegraphics[width=0.45\textwidth]{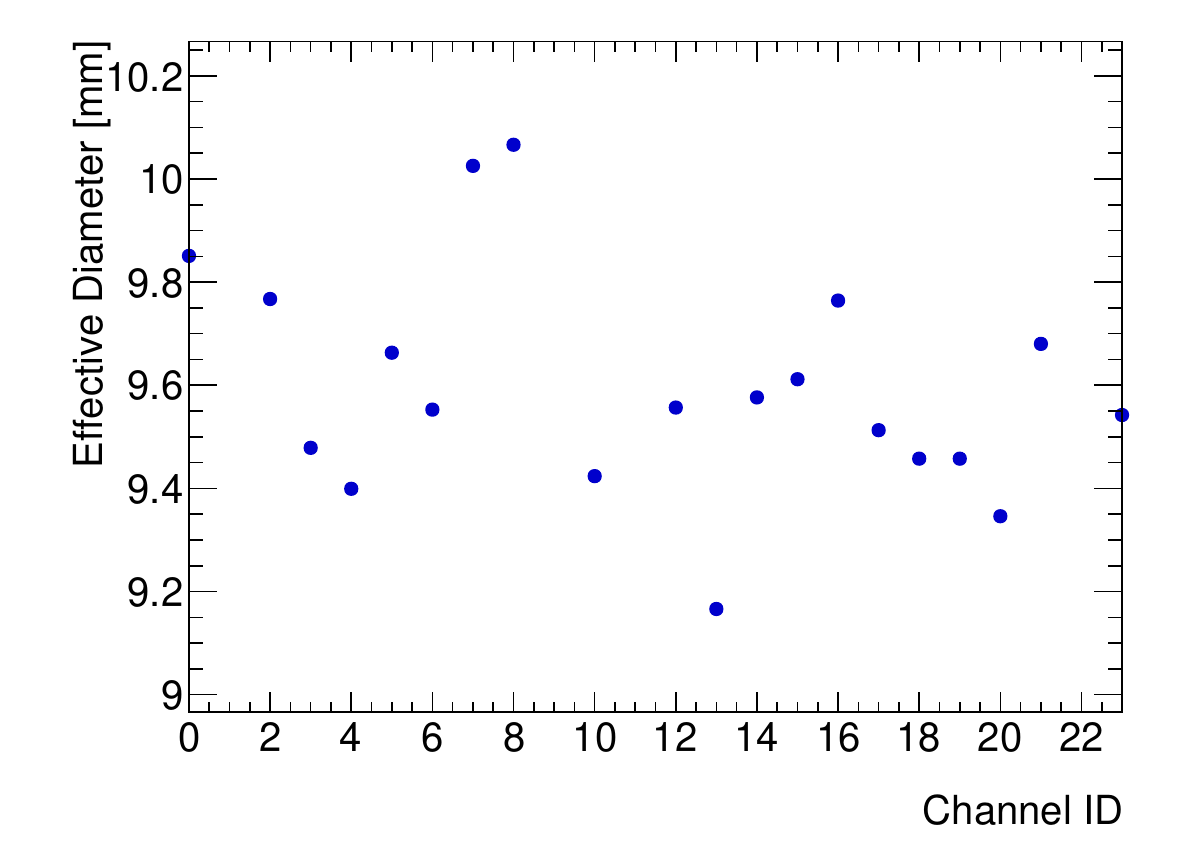} \label{fig:eff_length}} 
    \subfloat[]{\includegraphics[width=0.45\textwidth]{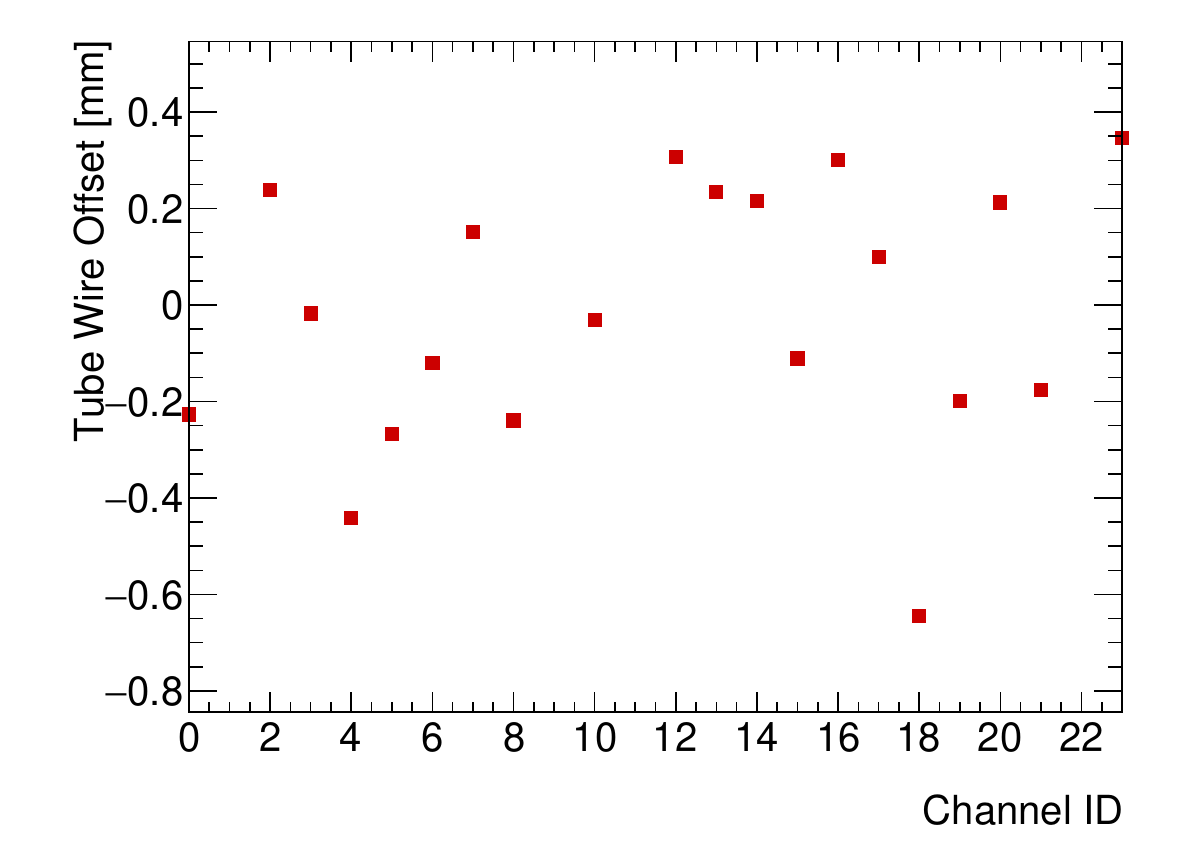} \label{fig:wire_offset}} 
    \caption{(a) Distribution of the effective diameter of 20 tubes; and (b) Distribution of the tube wire offset from the center of the tube.}
    \label{fig:tube wire and tube size}
\end{figure*}

 The asymmetry of the efficiency plot observed for a few straw tubes is attributed to the displacement of the anode wires relative to the tube center. We evaluated the anode wire offset and effective tube size for 20 tubes; the other three were excluded because low detection efficiencies near the tube edges prevented reliable edge fitting. We define an effective diameter as the region of the tube with a detection efficiency above 50\%, and this effective diameter is shown in Fig.~\ref{fig:eff_length}. On average, about 9.6 mm out of 10 mm have efficiencies above 50\%. Figure~\ref{fig:wire_offset} shows the wire offset for 20 tubes, while 3 tubes were located in the low efficiency region. The maximum wire offset was found to be about 0.7 mm for tube \#18, which has the most asymmetric efficiency shape shown in Fig.~\ref{fig:efficiency 2025}. 

\subsection{Spatial resolution along the tube direction}
\begin{figure*}[ht]
    \centering
    \subfloat[]{\includegraphics[width=0.45\textwidth]{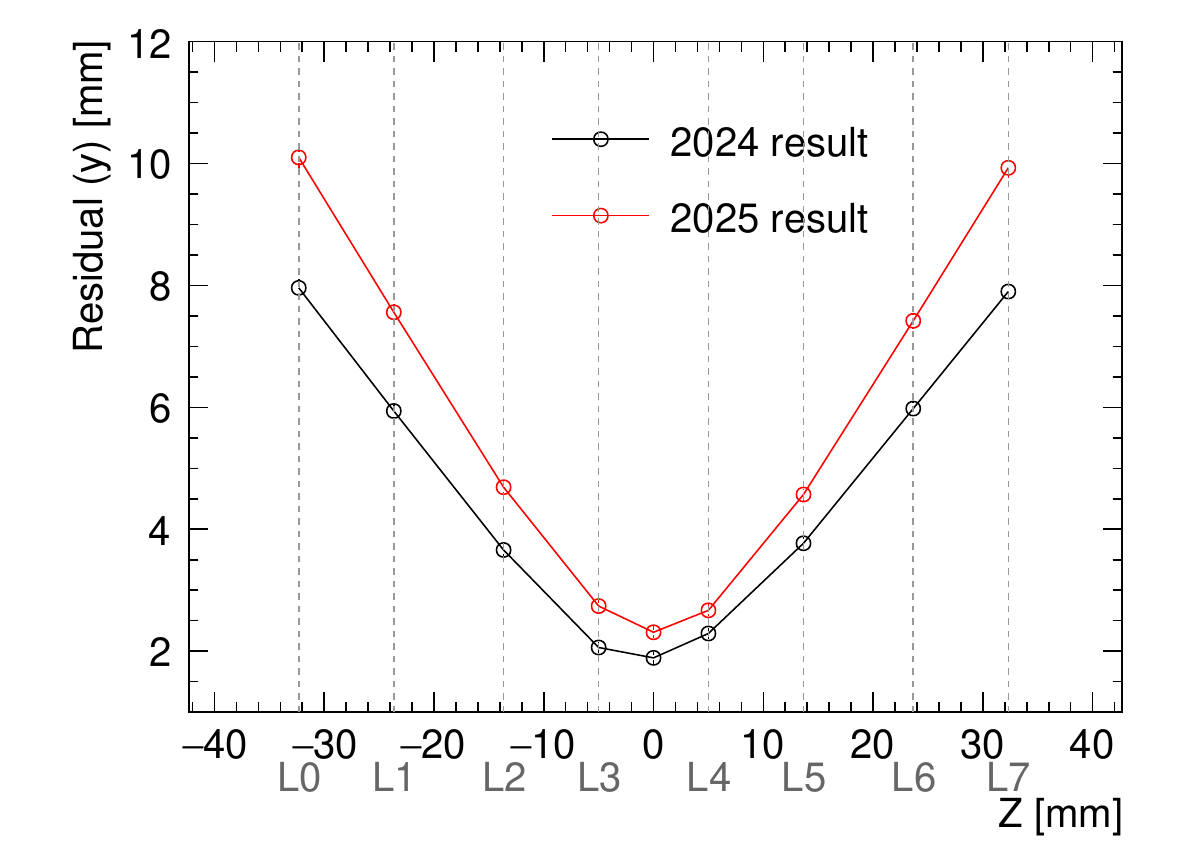} \label{fig:resolution_secondary}} 
    \subfloat[]{\includegraphics[width=0.45\textwidth]{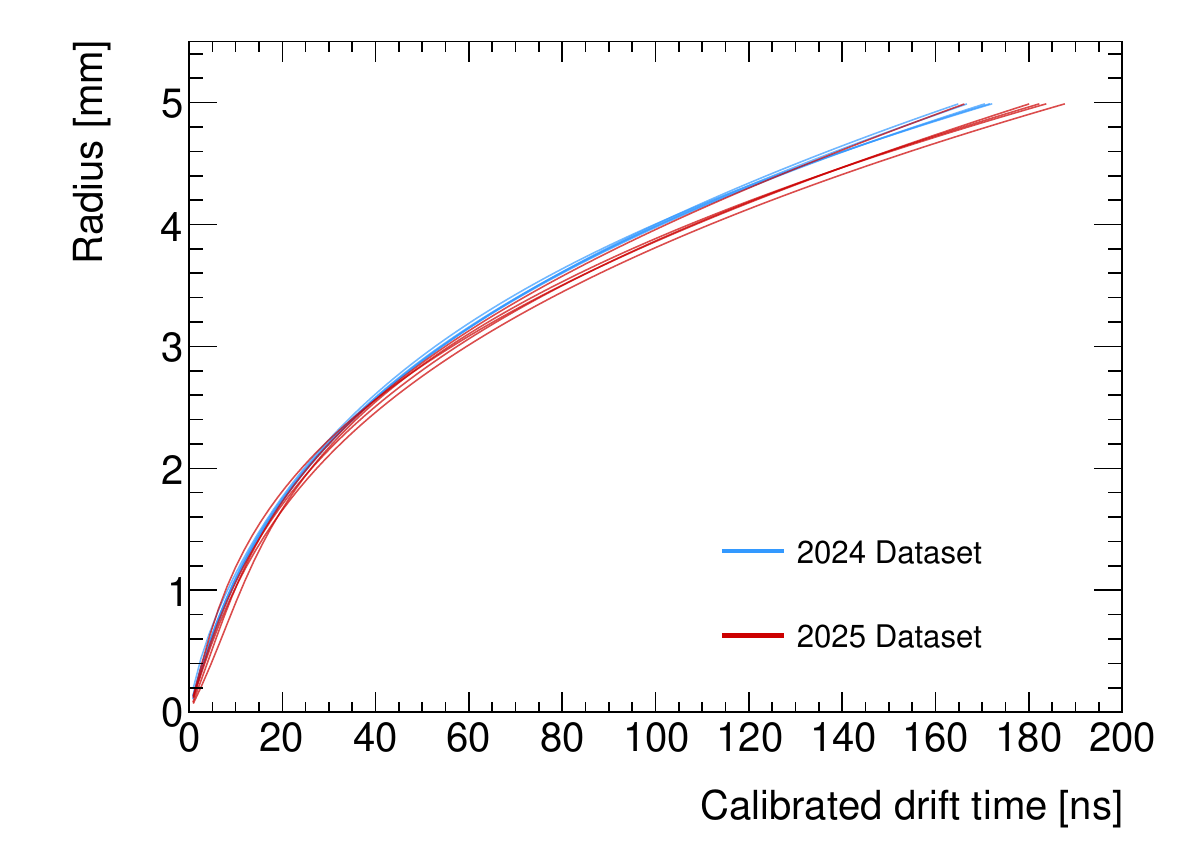}
    \label{fig:RT_curve}} 
    \caption{(a) The width of the secondary coordinate residual at different $z$ positions. Results from both 2024 and 2025 are shown; and (b) Comparison of the RT functions between the 2024 and 2025 datasets.}
    \label{fig:resolution vs z 2nd 2025}
\end{figure*}

The spatial resolution along the tube direction was evaluated for the 2025 run using the same procedure outlined in Sect.~\ref{sec:2nd 2024}. Figure~\ref{fig:resolution_secondary} illustrates the relationship between the residual width along the tube direction and the tube $z$ position for datasets collected in both 2024 and 2025. As the distance from the detector increases, the spatial resolution deteriorates, directly reflecting the angular resolution of the straw chamber. For the 2025 data, the resolution at the center of the chamber is 2.31 mm, which is slightly worse than the 1.89 mm observed in 2024. This degradation can primarily be attributed to larger variations in the RT relations among the tubes used in the 2025 dataset. Figure~\ref{fig:RT_curve} compares the calibrated RT functions obtained from the 2024 and 2025 datasets. In addition, increased mechanical misalignment of the tubes in the 2025 configuration contributed to the larger uncertainty observed.
The contribution from the misalignment was estimated by propagating the uncertainty of the stereo slope. For a nominal stereo angle of $2^\circ$ with $\delta(\tan\alpha)=0.0005$, the relative uncertainty in the reconstructed secondary coordinate is about 1.4\%. The 2025 beam covers about 120 mm along the tube direction; assuming a uniform distribution, this corresponds to an RMS spread of \SI{34.6}{mm}. The resulting contribution to the secondary-coordinate residual width is therefore estimated to be 0.5 mm. 

\section{Conclusions}
We have performed two test beam studies of straw tubes that could potentially be used for the FCC-ee straw tracker. Dedicated algorithms were developed to analyze the data and extract the single tube spatial resolution for the primary coordinate in the $r-\phi$ plane as well as the spatial resolution for the secondary coordinate along the tube direction within a straw chamber. In addition, detection efficiency was measured as a function of the extrapolated position for each tube. Data collected in both 2024 and 2025 were analyzed and compared, yielding consistent results. The average single-tube spatial resolution in the primary  $r-\phi$ coordinate is approximately 110 {\micro m},  while the average spatial resolution in the secondary coordinate along the tube direction is approximately 2 mm. Most tubes show a single-tube detection efficiency above 98\%.

\section{Acknowledgments}
The authors would like to thank the PS/SPS Physics Coordination and the EHN1 beam line and infrastructure teams, and the DRD1 Collaborations. This work was supported in part by U.S. Department of Energy Grant DE-SC0007859.


\end{document}